\documentclass[preprint2]{aastex}

\slugcomment{Resubmitted to Astrophysical Journal on 5/30/2013}

\shorttitle{Near-Infrared Imaging Polarimetry of RY Tau}
\shortauthors{Takami et al.}

\begin{document}



\title{High-Contrast Near-Infrared Imaging Polarimetry of the Protoplanetary Disk around RY Tau\footnote{Based on data collected at Subaru Telescope, which is operated by the National Astronomical Observatory of Japan.}}


\author{Michihiro Takami\altaffilmark{1}, Jennifer L. Karr\altaffilmark{1}, Jun Hashimoto\altaffilmark{2}, Hyosun Kim\altaffilmark{1}, 
John Wisnewski\altaffilmark{3}, 
Thomas Henning\altaffilmark{4}, 
Carol A. Grady\altaffilmark{5,6}, 
Ryo Kandori\altaffilmark{2},
Klaus W. Hodapp\altaffilmark{7},
Tomoyuki Kudo\altaffilmark{8},
Nobuhiko Kusakabe\altaffilmark{2},
Mei-Yin Chou\altaffilmark{1}, 
Yoichi Itoh\altaffilmark{9}, 
Munetake Momose\altaffilmark{10}, 
Satoshi Mayama\altaffilmark{11}, 
Thayne Currie\altaffilmark{12,6}, 
Katherine B. Follette\altaffilmark{13}, 
Jungmi Kwon\altaffilmark{2,14}, 
Lyu Abe\altaffilmark{15}, 
Wolfgang Brandner\altaffilmark{4}, 
Timothy D. Brandt\altaffilmark{16}, 
Joseph Carson\altaffilmark{17}, 
Sebastian E. Egner\altaffilmark{8}, 
Markus Feldt\altaffilmark{4}, 
Olivier Guyon\altaffilmark{8}, 
Yutaka Hayano\altaffilmark{8}, 
Masahiko Hayashi\altaffilmark{2}, 
Saeko Hayashi\altaffilmark{8}, 
Miki Ishii\altaffilmark{8}, 
Masanori Iye\altaffilmark{2}, 
Markus Janson\altaffilmark{16}, 
Gillian R. Knapp\altaffilmark{16}, 
Masayuki Kuzuhara\altaffilmark{18,8}, 
Michael W. McElwain\altaffilmark{6}, 
Taro Matsuo\altaffilmark{19}, 
Shoken Miyama\altaffilmark{2}, 
Jun-Ichi Morino\altaffilmark{2}, 
Amaya Moro-Martin\altaffilmark{20}, 
Tetsuo Nishimura\altaffilmark{8}, 
Tae-Soo Pyo\altaffilmark{8}, 
Eugene Serabyn\altaffilmark{21}, 
Hiroshi Suto\altaffilmark{2}, 
Ryuji Suzuki\altaffilmark{2}, 
Naruhisa Takato\altaffilmark{8}, 
Hiroshi Terada\altaffilmark{8}, 
Christian Thalmann\altaffilmark{22}, 
Daigo Tomono\altaffilmark{8}, 
Edwin L. Turner\altaffilmark{16,23}, 
Makoto Watanabe\altaffilmark{24}, 
Toru Yamada\altaffilmark{25}, 
Hideki Takami\altaffilmark{2}, 
Tomonori Usuda\altaffilmark{8}, 
Motohide Tamura\altaffilmark{2}
}

%
%


\altaffiltext{1}{Institute of Astronomy and Astrophysics, Academia Sinica.
P.O. Box 23-141, Taipei 10617, Taiwan, R.O.C.; hiro@asiaa.sinica.edu.tw}
\altaffiltext{2}{National Astronomical Observatory of Japan, 2-21-1 Osawa, Mitaka, Tokyo 181-8588, Japan}
\altaffiltext{3}{H.L. Dodge Department of Physics and Astronomy, University of Oklahoma, 440 W Brooks St Norman, OK 73019, USA}
\altaffiltext{4}{Max Planck Institute for Astronomy, Koenigstuhl 17, D-69117 Heidelberg, Germany}
\altaffiltext{5}{Eureka Scientific, 2452 Delmer Suite 100, Oakland CA 96402, USA}
\altaffiltext{6}{ExoPlanets and Stellar Astrophysics Laboratory, Code 667, Goddard Space Flight Center, Greenbelt, MD 20771, USA} 
\altaffiltext{7}{Institute for Astronomy, University of Hawaii, 640 North AÕohoku Place, Hilo, HI 96720, USA}
\altaffiltext{8}{Subaru Telescope, 650 North AÕohoku Place, Hilo, HI 96720, USA}
\altaffiltext{9}{Nishi-Harima Astronomical Observatory, Center for Astronomy, University of Hyogo, 407-2 Nishigaichi, Sayo, Sayo, Hyogo 679-5313, Japan} 
\altaffiltext{10}{College of Science, Ibaraki University, 2-1-1 Bunkyo, Mito, Ibaraki 310-8512} 
\altaffiltext{11}{The Center for the Promotion of Integrated Sciences, The Graduate University for Advanced Studies~(SOKENDAI), Shonan International Village, Hayama-cho, Miura-gun, Kanagawa 240-0193, Japan} 
\altaffiltext{12}{Department of Astronomy and Astrophysics, University of Toronto, Toronto, ON, Canada} 
\altaffiltext{13}{Steward Observatory, University of Arizona, 933 N Cherry Ave, Tucson AZ 85721} 
\altaffiltext{14}{Department of Astronomical Science, The Graduate University for Advanced Studies (SOKENDAI), 2-21-1 Osawa, Mitaka, Tokyo 181-8588, Japan} 
%
%
\altaffiltext{15}{Laboratoire Lagrange (UMR 7293), Universit\`{e} de Nice-Sophia Antipolis, CNRS,
Observatoire de la C\^{o}te d'Azur, 28 avenue Valrose, 06108 Nice Cedex 2, France} 
\altaffiltext{16}{Department of Astrophysical Sciences, Princeton University, Peyton Hall, Ivy Lane, Princeton, NJ 08544, USA}
\altaffiltext{17}{Department of Physics and Astronomy, College of Charleston, 58 Coming St., Charleston, SC 29424, USA} 
\altaffiltext{18}{Department of Earth and Planetary Science, The University of Tokyo, 7-3-1 Hongo, Bunkyo-ku, Tokyo, 113-0033, Japan} 
\altaffiltext{19}{Department of Astronomy, Kyoto University, Kitashirakawa-Oiwake-cho, Sakyo-ku, Kyoto, Kyoto 606-8502, Japan} 
\altaffiltext{20}{Department of Astrophysics, CAB-CSIC/INTA, 28850 Torrej—n de Ardoz, Madrid, Spain} 
\altaffiltext{21}{Jet Propulsion Laboratory, California Institute of Technology, Pasadena, CA, 91109, USA} 
\altaffiltext{22}{Astronomical Institute "Anton Pannekoek", University of Amsterdam,
Postbus 94249, 1090 GE, Amsterdam, The Netherlands} 
\altaffiltext{23}{Kavli Institute for the Physics and Mathematics of the Universe, The University of Tokyo, Kashiwa 277-8568, Japan} 
\altaffiltext{24}{Department of Cosmosciences, Hokkaido University, Kita-ku, Sapporo, Hokkaido 060-0810, Japan} 
\altaffiltext{25}{Astronomical Institute, Tohoku University, Aoba-ku, Sendai, Miyagi 980-8578, Japan} 


\begin{abstract}
We present near-infrared coronagraphic imaging polarimetry of RY Tau. The scattered light in the circumstellar environment was imaged at $H$-band at a high resolution ($\sim$0".05) for the first time, using Subaru-HiCIAO. The observed polarized intensity ($PI$) distribution shows a butterfly-like distribution of bright emission with an angular scale similar to the disk observed at millimeter wavelengths. This distribution is offset toward the blueshifted jet, indicating the presence of a geometrically thick disk or a remnant envelope, and therefore the earliest stage of the Class II evolutionary phase. We perform comparisons between the observed $PI$ distribution and disk models with: (1) full radiative transfer code, using the spectral energy distribution (SED) to constrain the disk parameters; and (2) monochromatic simulations of scattered light which explore a wide range of parameters space to constrain the disk and dust parameters. We show that these models cannot consistently explain the observed $PI$ distribution, SED, and the viewing angle inferred by millimeter interferometry. We suggest that the scattered light in the near-infrared is associated with an optically thin and geometrically thick layer above the disk surface, with the surface responsible for the infrared SED. 
Half of the scattered light and thermal radiation in this layer illuminates the disk surface, and this process may significantly affect the thermal structure of the disk.
\end{abstract}





\section{Introduction}
Optical and near-infrared observations have revealed structures in protoplanetary disks at the highest angular resolutions currently available. This has provided powerful tools for investigating the possibility of ongoing planet formation and to test the related theories. In particular, the technique of coronagraphic imaging has been extensively used to suppress the stellar flux and detect scattered light from the disk surface with high sensitivities \citep[see][for a review]{Watson07_PPV}. Without this technique, resolved images of protoplanetary disks can only be obtained in limited circumstences at optical and near-infrared wavelengths, via an edge-on view or silhouette against bright background nebular emission \citep[e.g.,][for a review]{McCaughrean00}.

Using coronagraphy, the ongoing survey ``Strategic Explorations of Exoplanets and Disks with Subaru" \citep[SEEDS,][]{Tamura09} with Subaru-HiCIAO \citep{Tamura06_HiCIAO} and AO188 \citep{Hayano04} has recently discovered structures in a number of protoplanetary disks \citep{Thalmann10}, in particular in the polarized intensity (hereafter $PI$) distribution in the near infrared \citep{Hashimoto11,Hashimoto12,Muto12,Kusakabe12,Tanii12,Mayama12,Dong12b,Grady13,Follette13}. $PI$ imaging has been used for observations of most of the disks in the SEEDS program, since this suffers significantly less contamination from the stellar flux than the normal intensity $I$.
Some disks are associated with spiral structures, disk holes, or azimuthal gaps in ring-like flux distributions, which are potential signatures of ongoing planet formation.
The scattered light from the disk is also useful for probing grain growth \citep[e.g.,][]{Min12,Tanii12} which may be related to formation of rocky cores.

The goal of the SEEDS program for protoplanetary disks is to observe a large number of objects with different stellar masses and ages and understand the evolution of disk structures and grain growth, and therefore the environment of possible ongoing planet formation. In this paper we present near-infrared coronagraphic imaging of RY Tau from the SEEDS program, the first publication of the near-infrared scattered light associated with the disk around this star. RY Tau is an active pre-main sequence star with a stellar mass of 2 $M_{\sun}$ \citep{Calvet04,Isella09}. The estimated age ranges from 0.5 \citep{Isella09} to 8 Myr \citep{Calvet04}. The star is associated with a relatively massive disk \citep[$3 \times 10^{-3} - 10^{-1}$ $M_{\sun}$,][]{Isella09,Isella10} with a large infrared excess at near-to-far infrared wavelengths \citep[e.g.,][]{Robitaille07}, an optical jet \citep{St-Onge08,Agra09}, a scattering nebulosity due to the remnant of the envelope \citep[e.g.,][]{Nakajima95,St-Onge08}, and a large time variation in optical photometry and spectroscopy \citep[e.g.,][]{Petrov99}.
These indicate a relatively young evolutionary phase. Despite this, the H$\alpha$ equivalent width observed over the last 30 years is relatively low \citep[8--20 \AA, see][and references therein]{Chou13}, similar to more evolved pre-main sequence stars.

Recent observations with millimeter interferometry seem to show evidence of a hole in the disk seen at 1.3-mm, with a radius of ~15 AU \citep{Isella10}. Such a hole can be made by tidal interaction between the inner disk and protoplanets \citep[e.g.,][]{Papaloizou07,Zhu11}. These disks are called Òtransitional disksÓ \citep[e.g.,][and references therein; Mayama et al. 2012]{Hughes09,Brown09}. However, RY Tau is different from the other transitional disks because it still drives a jet and has no evidence for a deficit of warm thermal emission near 10 $\micron$ \citep[e.g.,][]{Robitaille07}.


The rest of the paper is organized as follows.
In Section 2 we summarize our observations and data reduction. In Section 3 we show the observed $PI$ distribution and polarization.
We then perform comparisons with simulations of scattered light using conventional disk and dust models with the two approaches described below. In Section 4 we use the full radiative transfer tools developed by \citet{Whitney03a,Robitaille06,Robitaille07} with a disk geometry obtained by fitting the spectral energy distribution (SED). The model SEDs include the processes of scattering, absorption, and re-radiation on dust grains at all the wavelengths from UV to radio.
In Section 5 we use monochromatic simulations of scattered light with our own dedicated code to attempt to better fit the observed $PI$ distribution. Although the thermal structure and re-radiation process are not included, the emission from the disk surface is dominated by scattered light on dust grains, and this simplification allows us to conduct simulations covering a large parameter space in the disk geometry and grain size distribution.

In Section 6 we discuss the implications for the scattering geometry and dust grains, and the possible origins of the non-axisymmetry in the observed $PI$ distribution.
Throughout the paper we adopt a distance to the target of 140 pc \citep{Wichmann98}.


\section{Observations and Data Reduction}
Observations were made on 2011 January 27 using HiCIAO and AO 188 at Subaru 8.2-m.
As with several other SEEDS observations, the polarization was measured by rotating the half waveplate to four angular positions (in the order 0$^\circ$, 45$^\circ$, 22.5$^\circ$, and 67.5$^\circ$) using the PDI (Polarization Differential Imaging) +ADI (Angular Differential Imaging) mode. A single Wollaston prism was used to split incident light into two images, each with a 20" $\times$ 9" field of view and a pixel scale of 9.5 mas pixel$^{-1}$. We obtained 13 full waveplate rotation cycles, taking a 30-s exposure per waveplate position, with a 0".3-diameter coronagraphic mask.
The field rotation was $\sim$8$^\circ$ during the observations.

The data were reduced using the standard approach for polarimetric differential imaging \citep{Hinkley09} as well as the other SEEDS studies \citep[e.g.,][]{Hashimoto11,Hashimoto12,Kusakabe12}. The reduction was made using the Image Reduction and Analysis Facility (IRAF)\footnote{IRAF is distributed by National Optical Astronomy Observatory, which is operated by the Association of Universities for Research in Astronomy, Inc., under cooperative agreement with the National Science Foundation.}, pyRAF and python. Telescope and instrument polarization was corrected following \citet{Joos08}.

We find that the intensity ($I$) distribution observed in the disk region varies between exposures. This implies that the correction of the point-spread function with adaptive optics (AO) was not stable during the observations due to the moderate quality observing conditions, resulting in the stellar flux leaking out to a different degree in different images. We therefore use four images (at the waveplate positions of 0$^\circ$, 45$^\circ$, 22.5$^\circ$, and 67.5$^\circ$) with the minimum intensity distribution (i.e., minimum flux for the halo of the point-spread function associated with the star) to derive a lower limit for the degree of polarization. Even in these images, the $I$ distribution is centrosymmetric, i.e., very different from that expected for the disk associated with RY Tau (Sections 3--5), suggesting that they are severely contaminated by the stellar flux.

The unstable AO correction also resulted in different $PI$ distributions during the observations. To investigate this effect, we have calculated the co-added $PI$ images in a few different ways: i.e., taking an average or median for all the data set (i.e., 13 full waveplate rotation cycles), or selecting the best data sets and averaging them. All of the methods provide almost identical results.
In the rest of the paper, we use the median images for the 13 full waveplate rotation cycles, with a total integration time of 1560 s.

Before obtaining the coronagraphic frames we observed the object without a coronagraphic mask to measure the integrated stellar $I$ flux (hereafter $I_*$) and normalize the $PI$ flux for each pixel. We obtained three frames with a 1.5-s exposure,
with a half-wave plate P.A. of 0$^\circ$ and an 1-\% ND filter. The flux $I_*$ is obtained by integrating the $o$- and $e$-fluxes over the space in the same exposure. From these images and the individual science images we also estimate a median Strehl ratio of 0.30--0.37 during the RY Tau observations.

After observing RY Tau with the coronagraphic mask we observed a reference main sequence star (HD 282411) with three 10-s exposures, with a half-wave plate P.A. of 0$^\circ$ and a 0.1-\% ND filter, and without a coronagraphic mask. These images (average Strehl ratio of 0.56) will be used to convolve the simulated images in the later sections.

From the above images we derive an $H$-magnitude for RY Tau of 5.7. This is $\sim$0.4 mag brighter than the previous measurements of 6.1 mag by \citet{Kenyon95} and the 2MASS all-sky survey \citep[e.g., ][]{Robitaille07}. This discrepancy can be attributed to variability due to obscuration by circumstellar dust (Sections 6.3).


\section{Observed $PI$ Distribution}

\begin{figure}
\epsscale{1.1}
\plotone{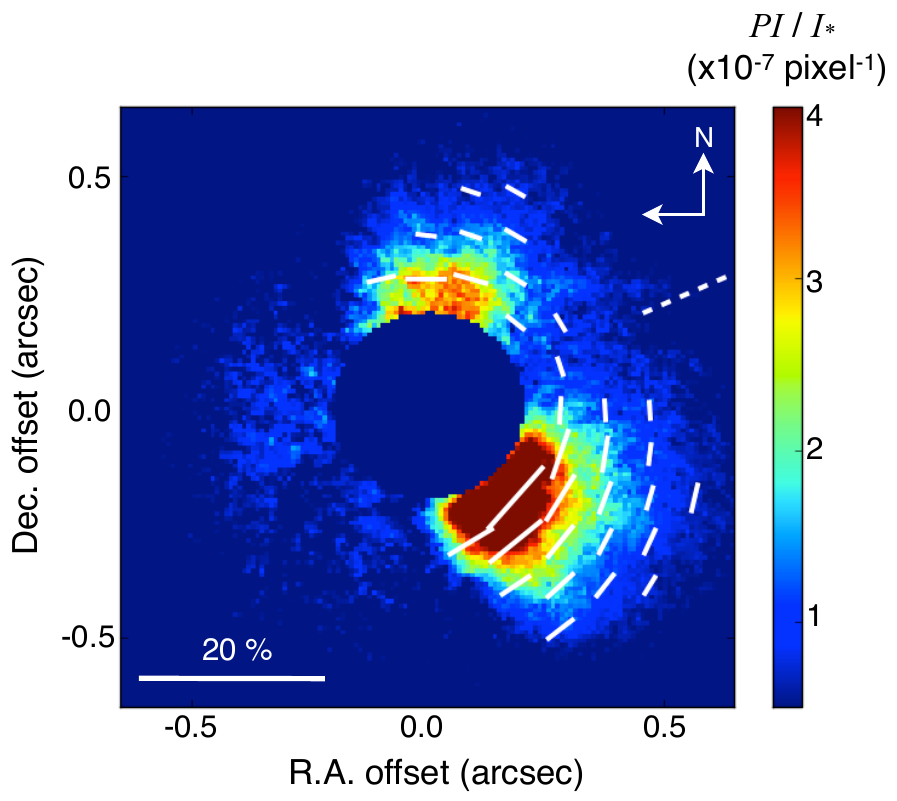}
\caption{Distribution of $PI$ flux and polarization vectors in $H$-band (1.65 $\micron$). The $PI$ flux at each pixel is normalized to the integrated stellar $I$ flux observed without the coronagraphic mask ($PI/I_* = 10^{-7}$ corresponds to 6.1 mJy arcsec$^{-2}$). In the $PI$ image we set a software aperture for the coronagraphic mask of 0".4 in diameter, slightly larger than that in the optics (0".3 in diameter) to show the distribution only where the measurements are reliable. The thin dashed line shows the direction of the extended jet observed by \citet{Agra09} (P.A.=294$^\circ$). The faint extended component in the northeast to south is probably due to an artifact due to unstable AO correction in the modest observing conditions. The degree of polarization shown here is a lower limit (see text). This is measured for individual 11$\times$11-pixel bins (corresponding to 0".1$\times$0".1), and shown for those in which the mean $PI/I_*$ is larger than $0.4 \times10^{-7}$. \label{fig1}}
\end{figure}

Figure \ref{fig1} shows the distribution of the $PI$ flux overlaid with the polarization vectors.
The $PI$ distribution is elongated along the major axis of the disk observed via millimeter interferometry \citep[P.A.=24$^\circ$/204$^\circ$,][]{Isella10}. Its angular scale is $\sim$1".0 ($\sim$140 AU) and $\sim$0".6 ($\sim$80 AU) along the major and minor axes, respectively. The angular scale for the major axis is similar to that observed by \citet{Isella10}.

The bright part of the $PI$ emission is offset from the star towards the blueshifted jet \citep[P.A.=294$^\circ$, ][]{St-Onge08,Agra09}, i.e., the far side of the disk. The $PI$ distribution shows a minimum along the direction of the jet (P.A.=294$^\circ$), and increases to maxima at P.A.s of $\sim$210$^\circ$ and $\sim$350$^\circ$. This butterfly-like morphology in the $PI$ distribution is similar to that modeled for some of the flared disks seen at optical wavelengths by \citet{Min12}. While the morphology in the $PI$ distribution is relatively symmetric about the jet (and disk) axis, its brightness is asymmetric. The southwest side is brighter than the northern side by a factor of $\sim$2 near the peaks ($PI/I_*\sim 8 \times 10^{-7}$ and $\sim 3 \times 10^{-7}$ per pixel, respectively). 



The observed $PI$ distribution, asymmetric with respect to the major axis of the disk, contrasts with the other disks observed in the SEEDS program \citep{Hashimoto11,Hashimoto12,Muto12,Kusakabe12,Tanii12,Mayama12,Grady13,Follette13}, in which the $PI$ distribution is generally symmetric. This indicates that RY Tau is associated with a geometrically thick disk or a remnant envelope, and therefore at a younger evolutionary stage than the others \citep[see, e.g., ][]{Fischer94,Whitney03b}.
This agrees with the stellar age of $\sim$0.5 Myr estimated by \citet{Isella10}, i.e., the star is in the earliest stage of the Class II phase, in which the star becomes visible at optical wavelengths \citep[e.g.,][]{Stahler05}.

Simulations show that such a disk is associated with faint emission at the other side of the disk
(Fisher et al. 1994; Whitney et al. 2003a; Section 5).
Although Figure \ref{fig1} shows a similar $PI$ distribution in the northeast to south of the star, this region is contaminated by an artifact caused by the variation of the halo in the point-spread function (PSF) of the star during the modest observing conditions. From our image we estimate an upper limit for $PI/I_*$ of $0.4 \times 10^{-7}$ per pixel.

The polarization vectors show a centrosymmetric pattern, as observed in several other disks. This is consistent with the scattered flux being dominated by a single scattering, with multiple scatterings being negligible (Sections 6.2). As described in Section 2, the degree of polarization shown here is a lower limit due to contamination from the  PSF halo of the star in the $I$ flux distribution. We measure a lower limit for the degree of polarization of $9 \pm 1$ \% at the peak in the southwest. The actual degree of polarization due to the scattered light would be significantly higher than this limit.


\section{Conventional Full-Radiative Transfer Models with Fitting Spectral Energy Distributions}

\citet{Robitaille07} developed a fitting tool for the SEDs of young stellar objects, and applied it to sources including RY Tau to investigate the physical parameters of the star, disk and envelope.
This fitting is made via comparison between the observed SED and a grid of 200,000 modeled SEDs covering a wide range of parameter space calculated by \citet{Robitaille06}.
Once we derive the parameters for the star, disk, and envelope using the above SED fitter, we can obtain the simulated $PI$ image of the scattered light in the disk and envelope using the \citet{Robitaille06} code.

In this section we use this approach to attempt to reproduce the observed $PI$ image. In Section 4.1 we briefly summarize the models based on the description in \citet{Robitaille06,Whitney03b,Whitney03a}, and our procedure for obtaining the simulated $PI$ images. In Section 4.2 we show the results of the SED fitting and the $PI$ image of the disks based on the fitted parameters.

Note that we use an axisymmetric distribution of the circumstellar material for the simulations in this and following sections. This implies that, in principle, these simulations cannot reproduce the asymmetry at the $PI$ distribution about the rotation axis of the disk (and the jet axis) described in Section 3. Simulations for an asymmetric $PI$ distribution are beyond the scope of this paper. The implications for this asymmetry is discussed in Section 6.3.

\subsection{\citet{Robitaille06,Robitaille07} Models}
The modeled system consists of a pre-main sequence star with an axisymmetric circumstellar disk, an infalling flattened envelope, and an outflow cavity.
The SED is determined using full radiative transfer in the disk and envelope, i.e., including absorption, scattering and re-radiation of light by dust grains in addition to the flux directly observed from the star. The heating sources of the disks are stellar radiation and viscosity in the disk.
The stellar spectrum is parameterized by the star's effective temperature and radius, via comparisons with modeled spectra for stellar photospheres \citep{Kurucz93,Brott05}.

The density distribution of a standard flared accretion disk \citep[e.g.,][]{Shakura73,Lynden-Bell74} is described in cylindrical coordinate ($r$,$z$) by:
\begin{equation}
\rho (r,z) = \rho_0 \left[1-\sqrt{\frac{R_*}{r}} \right] \left(\frac{R_*}{r} \right)^\alpha ~ \rm exp~ \left\{- \frac{1}{2} \left[\frac{\it z}{h}\rm \right]^2 \right\},
\end{equation}
where $\rho_0$ is a constant to scale the density; $R_*$ is the stellar radius; $\alpha$ is the radial density exponent; and $h$ is the disk scale height. The scale height $h$ increases with radius as $h = h_0 r^\beta$, where $\beta$ is the flaring power ($\beta > 0$). According to more detailed models of disk structures by \citet{Chiang97,DAlessio99b}, the scale height exponent varies with radius in the inner disk but follows a similar power law beyond a few AU \citep{Cotera01}.
%
%
In addition to the above parameters the minimum and maximum radii of the disk are included as free parameters. 
The gas-to-dust mass ratio is assumed to be 100.

Throughout the simulations \citet{Robitaille06,Robitaille07} assume $\alpha = \beta+1$. This yields the surface density distribution $\Sigma (r) \propto r^{-1}$, approximately agreeing with that inferred from millimeter interferometry for disks associated with many low-mass pre-main sequence stars \citep{Andrews09,Andrews10b}. This surface density power index may not be consistent with the one for the RY Tau disk seen at millimeter wavelengths by \citet{Isella10}.
Even so, models with the above assumption fit the observed SEDs well, as shown in Section 4.2.

The density structure for the envelope is given by \citet{Ulrich76,Terebey84}. The envelope is associated with a cavity whose shape 
varies as
$z \propto  r^{1.5}$.
\citet{Robitaille06} contains further details for the envelope and outflow cavity. Although the envelope mass, cavity density etc. will be calculated using the SED fitter, their contribution to the SEDs of optically visible pre-main sequence stars such as RY Tau will be negligible (Section 4.2).
%
%

The dust composition assumes a mixture of astronomical silicates and graphite in solar abundance without an ice coating. The following two grain size distributions are used based on \citet{Wood02b,Whitney03b,Whitney03a}: (1) the distribution in the denser regions of the disk ($m_{H_2} > 10^{10}$ cm$^{-3}$) where one would expect significant grain growth up to 1 mm; and (2) the distribution in the more diffuse regions ($m_{H_2} < 10^{10}$ cm$^{-3}$) in which the grain sizes are slightly larger than those in the diffuse interstellar matter. The grain size distribution for the former is described in \citet{Wood02b}, and that for the latter is similar to \citet{Kim94} \citep[hereafter KMH;][]{Whitney03b}. The former affects the thermal balance in the disk via radiative transfer, and is successful in fitting the SED of the HH 30 disk \citep{Wood02b,Whitney03b,Whitney03a}. The latter grains are responsible for scattered light on the disk surface (and in the envelope) at optical and near-infrared wavelengths. The optical constants for silicate and graphite are taken from \citet{Laor93}.

The 200,000 SEDs provided by  the \citet{Robitaille06} code include flux and polarization spectra for 250 wavelengths (from 0.01 to 5000 \micron). The authors used 20,000 parameter sets and computed the results for 10 viewing angles from face-on to edge-on at equal intervals in the cosine of the inclination. The SED fitter developed by \citet{Robitaille07} searches for best fitting SEDs using the minimum $\chi^2$ method to fit the observed fluxes at a range of wavelengths. We set a distance to the object of 140 pc, and an acceptable range for the visual extinction $A_V$ of 2.0--2.4, based on \citet{Calvet04}.
See \citet{Robitaille07} for further details of the fitting process.

To derive the star/disk/envelope parameters for RY Tau we used their SED fitter using the photometric data tabulated in \citet{Robitaille07} (Table \ref{tbl_observed_SED}). These parameters are used to model the $PI$ images with the \citet{Robitaille06} radiative transfer code. We used $10^7$ photons for each case. 
The resultant images for the Stokes parameters $Q$ and $U$ were convolved with the PSF of the reference star (Section 2) before obtaining the $PI$ image using $PI=\sqrt{Q^2+U^2}$. Then the PI flux is normalized to the stellar $I$ flux $I_*$, and scaled to match the pixel size of Subaru-HiCIAO.

\subsection{Results}
\begin{figure}
\epsscale{1}
\plotone{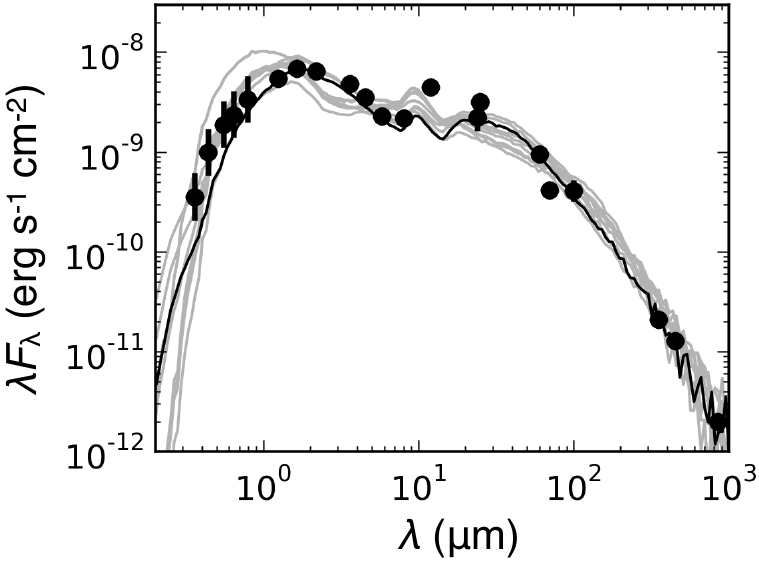}
\caption{Observed and modeled SEDs. Circles are the observed fluxes tabulated in \citet{Robitaille07}. { The error bars for the observed SED are shown only for those larger than the circles.} The solid curve shows the best-fit model using their online SED fitter. The gray curves are the same but for the next nine best fits. See Table \ref{best_fit_models} for the parameters for the star, disk, envelope, and inclination angle.
\label{SEDs}}
\end{figure}

Figure \ref{SEDs} shows the observed and ten best-fit SEDs. We note that the fluxes at different wavelengths were measured at different epochs (Table \ref{tbl_observed_SED}), and may be highly time-variable. In particular, \citet{Herbst94} reported a variation in the optical fluxes of a factor of $\sim$8 from 1961--1980 \citep[see also][for longer monitoring observations]{Petrov99}. Even so, the fluxes in Figure \ref{SEDs} are well fitted by a single SED except for the 12 $\micron$ flux observed using the InfRared Astronomical Satellite (IRAS). A larger excess at this wavelength may be attributed to bright silicate emission \citep{Honda06} in the filter coverage (8.5--15 $\micron$).

The parameters for the modeled SEDs are shown in Table \ref{best_fit_models} along with the $\chi^2$ value and the model ID specified by \citet{Robitaille06}. The constant used to scale the scale height relation ($h_0$) is fixed at $r$=50 AU from the star. The modeled SEDs are based on four sets of physical parameters for the star/disk/envelope with inclination angles of 57$^\circ$--76$^\circ$. The inclination angle for most of the SEDs and the disk mass for all the models are approximately consistent with those of \citet{Isella10} obtained using millimeter interferometry ($\ga 65^\circ$ and $\ga 3 \times 10^{-3} M_\sun$, respectively).

Despite the similarity in the shape of SEDs, some parameters are significantly different between models. The outer disk radius is $\sim$80 AU for two out of four models, but $\sim$120 and $\sim$400 AU for the remaining. The inner disk radius calculated with regard to the sublimation radius varies by a factor of $\sim$8 between models; the disk mass by a factor of $\sim$4; disk accretion rate by a factor of $\sim$20; and the envelope mass by a factor of $\sim 10^4$. The gas+dust density in the outflow cavity is $\sim 2 \times 10^{-21}$ g cm$^{-3}$ in two out of the four models, but zero for the remaining. 

\begin{figure*}
\epsscale{2.2}
\plotone{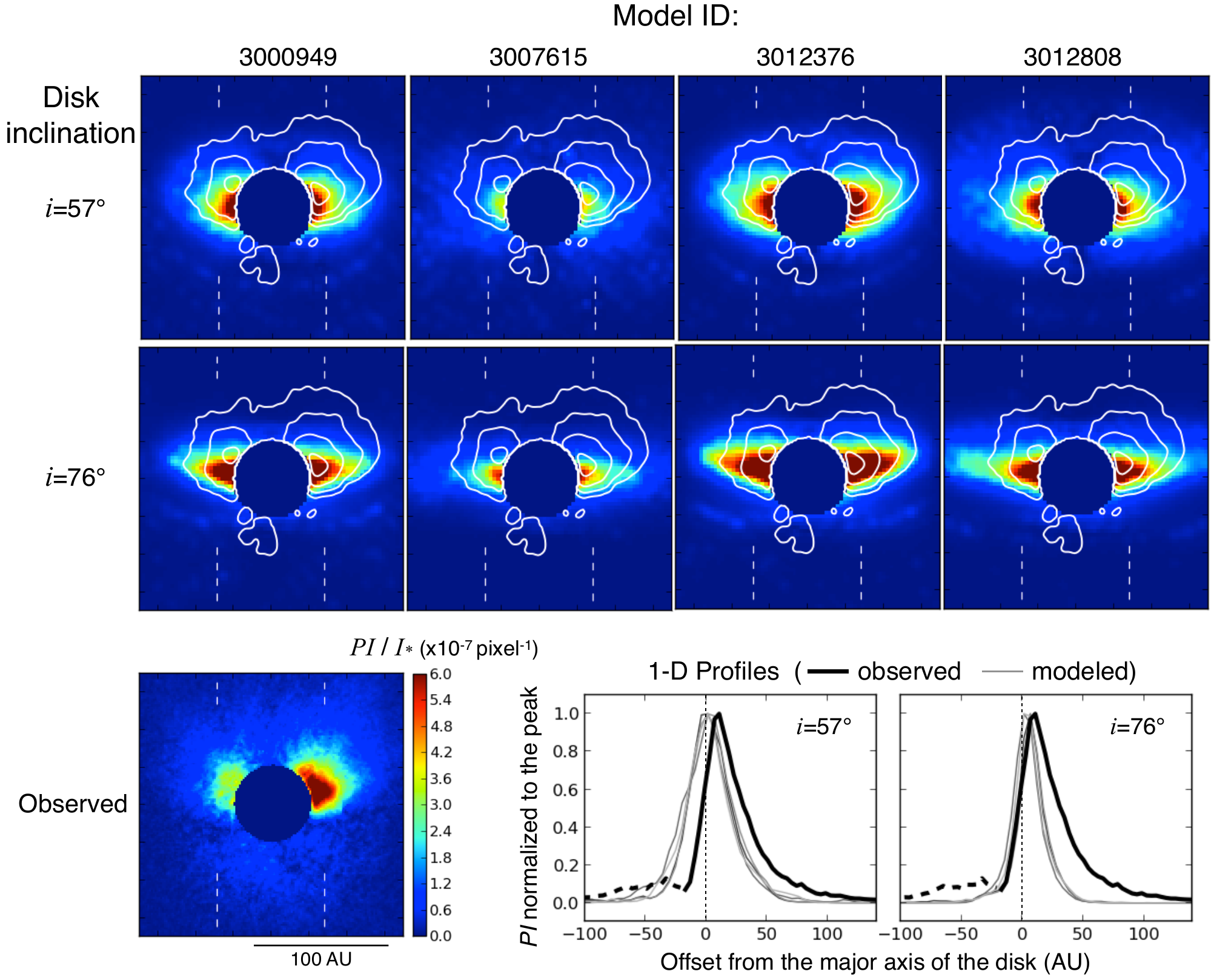}
\caption{Modeled $PI$ distributions for the best-fit SED models using the full radiative transfer code developed by \citet{Whitney03b,Whitney03a}. The observed $PI$ distribution is also shown at the bottom-left and in contour in the modeled images.
The color contrast is the same for all the images. The contour levels are 0.75, 1.5, 3, and 6$\times 10^{-7}$ per pixel scale of HiCIAO relative to the integrated stellar $I$ flux, and they are rotated by 72$^\circ$ from the observations. The dashed lines in the $PI$ images show the positions where we extract the 1-D distribution and show in the bottom-right. Each profile is made by averaging those at two sides, normalized to the peak $PI$ flux. The observed 1-D profile is dashed for the faint side of the disk where it is contaminated by an artifact (Section 3).
\label{PI_BW2008}}
\end{figure*}

Figure \ref{PI_BW2008} shows the simulated $PI$ images for the four physical parameter sets, with different inclination angles. Despite significant differences in the parameters described above, the four parameter sets result in $PI$ distributions that are strikingly similar to each other, except for the fact that model 3007615 produces a slightly fainter distribution than the others, due to the small flaring of the disk ($h_{\rm 50AU}=1.6$ AU, $\beta=1.12$).
This similarity between models, both for the SEDs and $PI$ distribution, can be attributed to the fact that both the infrared radiation and the scattered light at the disk surface are determined by the surface geometry of the disk.
The $PI$ distribution shown in Figure \ref{PI_BW2008} is dominated by scattering on the disk surface, and the contribution from the envelope and outflow cavity is significantly smaller. To investigate this, we removed the envelope 
from the models with a relatively massive envelope and diffuse dust layer in the cavity (models 3000949 and 3012376 according to the SED fitter), and ran the simulations again. We find that the $PI$ distribution is almost identical to the model with the envelope and cavity, in particular within $\sim$100 AU of the star.


The models at inclination angles $i$= 57$^\circ$ and 76$^\circ$ show two separated bright regions
similar to the observations. Furthermore, the modeled $PI$ flux normalized to the stellar $I$ flux ($PI/I_*$) is of the same order as the HiCIAO observations. 
However, the offset from the major axis of the disk is significantly smaller for all of the modeled $PI$ images compared to the observations. These offsets for the models and observations are clearly shown in the bottom right plots of Figure \ref{PI_BW2008}. While that of the observed image is clearly offset from the major axis, the modeled images are more symmetric about the major axis of the disk. This indicates that the disks inferred from the SEDs are geometrically thinner than that producing the observations\footnote{The ``thickness" here does not imply the scale height in Equation (1) but that between the midplane and disk surface determined with $\tau \sim 1$ from the star \citep[e.g.,][see also Section 5.1 and Figure \ref{Sprout_examples}]{Watson07_PPV}}, if the shape of the emission is attributed to the disk.



\section{Monochromatic Simulations with Conventional Disk Models}
In order to explore the effect of geometry, we relax the density distribution of the disk constrained by the SED fitting and focus on reproducing the $PI$ distribution. To do this we carry out monochromatic radiative transfer calculations using Monte-Carlo code developed by us  specifically for use with SEEDS observations (the Sprout code). We perform simulations over a large region of parameter space for the disk geometry, and also investigate the effects of different grain size distributions. We describe the details of the simulations in Section 5.1, and the results in Section 5.2.

In this section we focus on comparisons between the models and observations for the bright part of the $PI$ distribution. Although the modeled $PI$ distribution is associated with faint emission at the other side of the disk, we will not discuss this component. For this component one could easily reproduce a $PI$ flux consistent with observations (i.e., the upper limit), by adding extinction by a flattened envelope not included in the models in this section.

\subsection{Models and Simulations}

We follow the method described in \citet{Fischer94}. We place a central unpolarized light source equivalent to the star as the starting point for calculating the scattering of photons from the disk.
The light path for the next scattering position is calculated for an opacity distribution based on the disk in Equation (1) and the dust opacity described below. The scattering angle and Stokes parameters after scattering are calculated based on Mie theory. The Stokes parameters for each photon are initially set to ($I_0$,$Q_0$,$U_0$,$V_0$)=(1,0,0,0) and normalized to $\it I_{\rm out} \it = albedo \cdot \it I_{\rm in}$ after each scattering.

We use $10^6$ photons for each simulation.
The photons escaping from the disk are collected in imaginary detectors at different viewing angles.
For our purpose unscattered stellar photons are not collected. In order to normalize the $PI$ flux to the stellar $I$ flux, we separately calculate the expected number of photons for each viewing angle based on the incident number of photons and extinction.

\begin{figure}
\epsscale{1.1}
\plotone{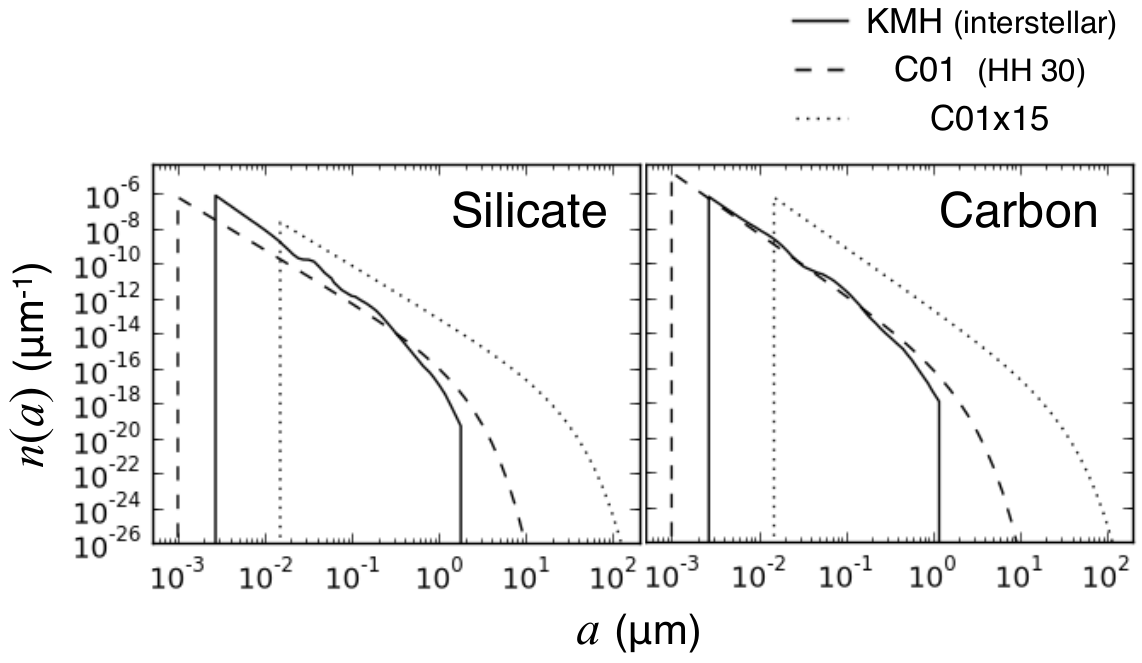}
\caption{
Size distributions of silicate and carbon grains for 
(1) interstellar medium \citep[KMH, ][]{Kim94},
(2) that used for explaining the scattered light in the HH 30 disk \citep[C01][]{Cotera01,Wood02b}, and
(3) that 15 times larger than (2) (C01$\times$15). \label{size_distributions}}
\end{figure}

For dust grains we use homogeneous spherical particles as commonly used in other studies, including \citet{Whitney92,Whitney03b,Whitney03a,Cotera01,Wood02b,Robitaille06}. 
We use
the grain compositions of \citet{Robitaille06,Robitaille07,Cotera01,Wood02b}, i.e., astronomical silicate and carbon dust without an ice coating. For the size distribution, we use (1) that of interstellar dust measured by KMH ($R_V=3.1$); (2) the larger size distributions used by \citet{Cotera01} and \citet{Wood02b} to reproduce the scattered light observed in the HH 30 disk (hereafter C01); and (3) same as C01 but with the grain size scaled by a factor of 15, preserving the total dust mass (hereafter C01$\times$15) (Figure \ref{size_distributions}). Note that the C01$\times$15 distribution would be too large for the disk surface of RY Tau. \citet{Honda06} conducted mid-infrared spectroscopy and analysis of silicate emission toward a number of pre-main sequence stars. These authors fit the spectrum of individual stars using models with two dust sizes (0.1 and 1.5 $\micron$), and showed that the disk surface of RY Tau has one of the smallest ratios of large-size grains to small-size grains of their sample (34 \%). This suggests that the grain size distribution in RY Tau is fairly close to that of the HH 30 disk, which has a central star that is significantly younger than most optically visible pre-main sequence stars \citep[e.g.,][]{Burrows96,Watson07_HH30}. However, we still show the results obtained with C01$\times$15 to investigate how the results are affected by the assumed size distribution.

Different authors use different types of carbon dust, either graphite or amorphous carbon \citep{Cotera01,Wood02b}. While graphite has been extensively used \citep[e.g.,][]{Draine84,Laor93,Kim94,Whitney03a,Robitaille06,Dong12b,Dong12a}, far-infrared SEDs of young stellar objects and evolved stars suggest the absence of graphite and presence of amorphous carbon in circumstellar dust \citep[][and references therein]{Jager98}. We still use graphite for the KHM distribution for consistency with the authors as their size distribution is determined assuming graphite for the carbon dust. We use amorphous carbon for C01 and C01$\times$15 for the above reason and following \citet{Cotera01,Wood02b}. 
The use of different carbon dusts does not significantly affect the modeled $PI$ distribution, and it does not affect the conclusions of the paper (Appendix A).

Calculations for Mie scattering are made using the code developed by \citet{Wiscombe96}. The optical constants for astronomical silicate and graphite are obtained from \citet{Draine84}: they are identical in the near infrared with the \citet{Laor93} values used for the simulations in Section 4. For amorphous carbon we use the optical constants of \citet{Jager98} with a pyrolysis temperature of 600$^\circ$C. As for the use of different carbon materials (amorphous carbon or graphite), amorphous carbon with different pyrolysis temperatures does not significantly affect the modeled $PI$ distribution and conclusions of the paper (Appendix A).

Table \ref{tbl_dust_properties} shows the physical properties (opacity, albedo, forward throwing parameter and the maximum degree of polarization) for the above three dust models.
%
Figure \ref{scat_properties} shows the scattering properties of these dust models at 1.65 \micron, i.e., $I$, $PI$, and the degree of polarization as a function of scattering angle, in the case where the incident light is not polarized. The $I$ and $PI$ fluxes for each scattering angle are derived by normalizing the scattering matrix elements $S_{11}$ and $-S_{12}$ by a constant $I_0$ to match the results to the ``weighted photon method" described above (i.e., $\int I/I_0 d\Omega =albedo$). The forward scattering is more significant for large grains models, but different dust models show a relatively similar distribution of scattered intensity $I/I_0$ at scattering angles of 40--180$^\circ$. The polarization shows a similar dependency on scattering angle, peaking at $\sim 90^\circ$, but the absolute value for the C01 and C01$\times$15 models are 30--40 \% lower than for the KMH. The $PI$ flux shows a maximum at scattering angles of $\theta$=$75^\circ$, $60^\circ$, and $28^\circ$ for the KMH, C01, and C01$\times$15 models, respectively, and decreases towards both sides for all these dust models.

We assume an outer radius for the disk of 80 AU, based on our HiCIAO observations and millimeter interferometry by \citet{Isella10}. The inner radius is set to 1 AU. We find that our results are not significantly affected by the choice of inner radius as long as it is within the coronagraphic mask ($r \sim$30 AU at a distance to the target of 140 pc) and the disk surface has a flared geometry. Note that the millimeter interferometry of \citet{Isella10} suggests the presence of a disk hole with a radius of $\sim$15 AU, while the presence of near-infrared emission suggests the presence of warm dust within 1 AU \citep{Akeson05,Pott10}.

The remaining free parameters for the density distribution of the disk are $\rho_0$, $h_{\rm 50AU}$, $\beta$, and $\alpha$.  The first three parameters all have a large effect on the offset of the $PI$ distribution from the major axis of the disk. This makes it difficult to search for the parameter sets which best fit the results. To overcome this problem, we set the total optical thickness between the star and the edge of the disk to 1 at a direction $\theta$ from the midplane of the disk. This angle will be adjusted to fit the degree of the offset of the $PI$ distribution from the major axis of the disk (see Figure \ref{Sprout_examples} for an example). This determines the parameter $\rho_0$, and as a result, our parameter searches will be made with the remaining two free parameters, i.e.,  $h_{\rm 50AU}$ and $\beta$. For the simulation in Section 5.2 we adopt $\alpha=\beta+1$ as we did in Section 4. The use of an independent $\alpha$ produces almost identical results.

In addition to $\rho_0$, $h_{\rm 50AU}$, $\beta$, and $\alpha$, Equation (1) also includes the stellar radius $R_*$, but its uncertainty does not affect the results of simulations. The stellar radius is significantly smaller than the radii for the density distribution we use (i.e., $>$1 AU), and therefore the term $\sqrt{R_*/r}$ produces only a very minor contribution to Equation (1). The equation also includes the term $(R_*/r)^\alpha$, but the stellar radius here can be replaced by any radius, depending on the radius at which we define $\rho_0$. Again, this parameter is not explicitly provided but scaled using the constraint of the optical thickness described above.

Throughout the simulations we assume that re-radiation from the disk is negligible, or that it occurs in very close proximity to the star so that the effect of the extended structure in the emission region far from the star is negligible. To prove the validity of this assumption, we ran this simulation code with the disk parameters in Table \ref{best_fit_models}, and found that the code reproduces the $PI$ distribution in Figure \ref{PI_BW2008} well.

\subsection{Results}

Figure \ref{Sprout_30deg} shows examples for 
the $PI$ distribution using the C01 grain size distribution (i.e., that of the HH 30 disk). The combination of the angle for $\tau = 1$ ($\theta = 30^\circ$) and the viewing angle (49$^\circ$ from the face-on view) are selected to approximately fit the observed morphology. 
As for the observations and models in Sections 3 and 4, the $PI$ flux displayed is normalized to the stellar $I$ flux.
The image convolution is made in the same manner as Section 4.
The contrast for each figure is adjusted in each image for the best morphological comparison for the bright part.

In Figure \ref{Sprout_30deg} all except the case of $h_{\rm 50AU}$= 25 AU, $\beta$=1.3 show a butterfly-like morphology similar to the observed $PI$ distribution. The peak $PI$ flux in these models, however, is 3--6 times larger than the observed value. Increasing $\beta$ produces a brighter $PI$ distribution. 
In the case of $\beta$=2.0 and 2.7, the modeled $PI/I_*$ is larger for $h_{\rm 50AU}$=25 AU than  $h_{\rm 50AU}$=5 AU by a factor of $\sim 1.5$. This difference is attributed to the fact that, for the latter model, the star suffers extinction from the edge of the disk which decreases $I_*$.
In the case of $\beta$=2.0 and 2.7, increasing $h_{\rm 50AU}$ also produces a larger $PI$ flux at the other side of the disk. 
The $PI$ distribution is fairly centrosymmetric for $h_{\rm 50AU}$= 25 AU and $\beta$=1.3 for any contrast.

In the same figure we also show the density profiles for the individual disks. The dashed, solid, and dotted curves show the positions at which $\tau$=0.5, 1, and 2, respectively, measured from the star. These curves indicate that the scattering layer is geometrically thin at the disk surface. Increasing $\beta$ produces a large flaring angle in the outer disk, which explains the brighter $PI$ flux as the surface is more easily illuminated. The curvature of the surface remains similar for the same $\beta$, however, the disk is not flared for $\beta$=1.3 and $h_{\rm 50AU}$. In this case, the scattering is dominated by the inner part of the disk, producing a centrosymmetric $PI$ distribution as described above.

The viewing angle selected for Figure \ref{Sprout_30deg} (49$^\circ$) is smaller than that determined from millimeter emission by \citet{Isella10} ($>65 ^\circ$). A larger viewing angle  does not reproduce the observed $PI$ morphology. Figure \ref{Sprout_25deg} shows examples for a viewing angle of 63$^\circ$. The disk parameters are identical to those for Figure \ref{Sprout_30deg}, but the optical thickness along the radial direction has been set to $\tau = 1$ at $\theta = 25^\circ$ to fit the offset of the bright $PI$ emission from the major disk axis. The results are similar to those of Figure \ref{Sprout_30deg}, but the bright part shows a thinner distribution in the vertical direction than the observations. Furthermore, a large $\beta$ results in a larger spatial extension at the bright side. To clearly show the former discrepancy, we extract the 1-D profiles at the positions indicated in Figures \ref{Sprout_30deg} and \ref{Sprout_25deg} and show them in Figure \ref{Sprout_profs}. The peak $PI$ fluxes are comparable to or larger than those in Figure \ref{Sprout_30deg}, and therefore significantly larger than the observations.

Figure \ref{Sprout_different_size_distributions} shows the $PI$ images for the same disk parameters as Figures \ref{Sprout_30deg} and \ref{Sprout_25deg} but with different dust models. Changing the dust model results in little variation in the $PI$ images. This can be explained by the fact that the $PI$ flux at the bright side is dominated by a single scattering with scattering angles $60^\circ -135^\circ$, and the $PI$ flux for this range of angles is similar for the three different dust models, as shown in Figure \ref{scat_properties}. In particular, the $PI$ fluxes shown in Figure \ref{scat_properties} for these scattering angles explain the fact that the $PI$ flux for the KMH model is larger by a factor of $\sim$2 than the others in Figure \ref{Sprout_different_size_distributions}. Table \ref{max_PI} shows the peak $PI/I_*$ for individual models. These are larger than the observations by a factor of 2--12 at the peak. 

In summary, the modeled morphology in the $PI$ distribution does not match the observations with the viewing angle inferred from the millimeter observations ($> 65^\circ$). As shown in Figures \ref{Sprout_30deg}, \ref{Sprout_25deg}, and \ref{Sprout_different_size_distributions}, the modeled $PI$ distribution from the bright side of the disk does not vary significantly with the value of $h_{\rm 50AU}$, $\beta$, or the choice of dust model once we determine the outer radius of the disk and the optical thickness in a given direction.
%
%
Some models may also be excluded due to the degree of polarization, the inferred disk mass or extinction toward the star. However, these do not provide constraints as clearly as those from the $PI$ flux. We discuss these constraints in Appendix B.
%


\section{Discussion}

In Section 6.1 we discuss the possible presence of an optically thin scattering layer above the disk which could be responsible for the observed $PI$ distribution. In Section 6.2 we briefly discuss whether the use of more realistic dust models might still fit the observations using a conventional disk model. In Section 6.3 we discuss the implications of the asymmetry of the $PI$ distribution about the jet/disk axis that was shown in Section 3.

\subsection{Geometry of the Scattering Layer}

In Sections 4 and 5, we found inconsistencies between the models obtained with conventional disk and dust models and the observed $PI$ distribution. The disk geometry obtained from the SED corresponds to a $PI$ distribution with a significantly smaller offset from the major axis of the disk (Section 4). On the other hand, the disk geometries which reproduce this offset require a smaller viewing angle (i.e., close to face-on) than that determined through millimeter observations (Section 5). 
We have not explored the full parameter space, but once we fix the parameters to reproduce the angular scale along the major axis and the offset of the bright $PI$ emission on the far side of the disk, varying disk and dust parameters
has little effect on the resulting $PI$ distribution.
%

The results of the two sets of simulations (full radiative transfer and monochromatic scattering) may imply that the system consists of (1) a geometrically thin disk which is partially responsible for the infrared SED but does not contribute to the $PI$ flux in the near infrared; and (2) an optically thin and geometrically thick upper layer which is responsible for the observed $PI$ distribution in the near infrared and the remaining mid-to-far infrared flux. 
A similar geometry has been proposed for SEDs observed for some Herbig AeBe stars \citep[][Group I]{Meeus01}. \citet{Follette13} also discuss such a geometry for the SR 21 disk in which the presence of a inner cavity is indicated by sub-millimeter observations but not seen in the scattered light in near-infrared.

Figure \ref{disk_sketches} shows a schematic view for the disk, an optically thin layer, stellar radiation, scattered light and infrared re-radiation. A rough analytic estimate for the optical thickness, density, and mass of such a layer is described in Appendix C.
The observed $PI$ image does not show evidence for emission along the major axis without an offset as modeled in Section 4. This can be explained if 
the thickness of the disk is significantly smaller than those used in Section 4, which provide a $PI$ flux level comparable to the observations.

Such a scattering geometry can easily explain the observed $PI$ flux and offset from the major axis of the disk, by adjusting the column density and vertical distribution of the optically thin layer, respectively. Here we qualitatively demonstrate this using the Sprout code with an optically thin upper layer using Equation (1) and an optically thick disk with a geometrical thickness of zero. To focus on the scattered light outside the coronagraphic mask, we set the density to be zero for disk radii within the mask ($r < 28$ AU).

Figure \ref{Sprout_opt_thin} shows an example of simulations using the KMH dust model, with a viewing angle of 70$^\circ$. The butterfly-like bright $PI$ distribution approximately matches the observations. We measure the modeled polarization degree to be 25 \% at a position corresponding to the $PI$ peak of the observations. This is also consistent with the observations of $\gg$9 \% measured in Section 3. The faint extended emission at the other side of the disk is brighter than the upper limit of the observations by a factor of $\sim$2. The modeled flux in this region could be adjusted, e.g., if we included extinction by a remnant envelope, or if we define an exponential cutoff for the outer radii of the optically thin layers like that used to reproduce the millimeter flux distribution of the disk \citep[e.g.,][]{Andrews09,Isella10}.

The use of different dust models (C01, C01$\times$15) provides similar results to Figure \ref{Sprout_opt_thin}, but the polarization degree is significantly smaller: 13--14 \% at the position corresponding to the $PI$ peak of the observations. Such a polarization may be too close to the lower limit of the observations of $\sim$9 \%. It is beyond the scope of the paper to identify the best dust models using the modeled degree of polarization. See Section 6.2 for more complex and realistic dust models than used in our study.

Use of a geometrically thinner disk than Section 4 implies that the disk gets less stellar flux, thereby resulting in less infrared excess than that shown in Figure \ref{SEDs}. This issue could be solved as follows.
As suggested by \citet{Chiang97}, an optically thin layer on the disk surface scatters or reemits directly to space about half the radiation, while the other half is scattered/emitted inward. The extra illumination from the optically thinner layer would warm up the disk and enhance the observed infrared flux. 
Although such a optically thin layer is usually assumed to be geometrically thin \citep[e.g.,][]{Chiang97,Dullemond07_PPV}, the same physical mechanism should work even in the case where the layer is geometrically thick.
If the scattered light is dominated by the optically thin layer, this implies that the disk surface would be illuminated more significantly from this layer than the direct stellar flux. Therefore, the presence of this layer may significantly affect the temperature structure of the disk.

In addition, the thermal radiation from the optically thin layer may directly contribute to the infrared SED \citep[e.g.,][]{Chiang97}. However, the fraction compared to the disk flux would be significantly smaller than that of the scattered light in the near-infrared due to significantly smaller dust opacities in the mid-to-far infrared \citep[e.g.,][]{Wood02b,Whitney03b,Whitney03a,Dong12a}.

\subsection{Dust Grains}

While homogeneous spherical particles are widely used for the dust in disks \citep[][]{Whitney92,Fischer94,Whitney03b,Whitney03a,Robitaille06,Robitaille07}, dust particles in practice appear to be aggregated \citep[e.g.,][]{Henning96,Dominik09}, and may also be coated with ice \citep{Malfait99,Meeus01,Honda09}. These facts would produce different optical properties.
\citet{Min12} conducted simulations for scattered light from the disk using both spherical and aggregated particles, and show that a lower $PI$ flux is expected for the latter due to the extremely large forward scattering. This would allow us to better match the $PI$ flux we simulated in Section 5 for the disk surface, as our values were significantly larger than the observations for many cases.

However, the disks used in Section 5 would produce significantly different SEDs than those shown in Figure \ref{SEDs}. A large flaring allows the disk to receive more stellar photons, which warms up the disk more efficiently, thereby producing a larger excess than the thinner disks we used in Section 4. Furthermore, some disks used in Section 5 have a large extinction toward the star (Appendix B). This would result in a double peaked SED in the optical to near-infrared, and far infrared, respectively \citep[e.g.,][]{DAlessio99,Robitaille07}. The former problem could be overcome if the dust particles had a smaller emissivity in mid-to-far infrared, or large albedo at UV-to-IR in order to impede the absorption of the stellar photons.
To investigate further, one would need a detailed understanding of the optical properties of the aggregates over a wide range of wavelengths, plus simulations with full radiative transfer.

Alternatively, observations of the $PI$ distribution at different wavelengths would give useful constraints for the nature of the dust grains. We expect that the $PI$ flux is dominated by the photons with a single scattering, even in the case of optically thick disks (such as those used in Sections 4 and 5) in which multiple scattering occurs as well as single scattering.
The contribution from multiple scattering would be negligible for the reasons below.
In the case of small grains (e.g., KMH), where photons are fairly isotropically scattered, the scattered photons have polarizations with a variety of P.A., canceling each other out.
In the case of large grains (e.g., C01 and C01$\times$15), in which most of the photons are scattered forward, the polarization of individual photons is significantly reduced after the first scattering.

To demonstrate this, we re-ran the simulations for the disks in Section 5, removing all the photons which experience multiple scatterings. We find that the results are almost identical for the $PI$ flux distribution, with a flux difference within 10 \% at the peak. This is in contrast with the $I$ flux for which multiple scatterings are responsible for 20--40 \% of the flux. Throughout, studies of the $PI$ flux allow a more accurate investigation of the dust properties than studies of the $I$ flux, in particular for optically thick disks.

\subsection{Implication of the Asymmetric $PI$ Distribution}
%

As shown in Section 3, the observed $PI$ distribution is asymmetric about the rotation axis of the disk: the southwest side is brighter than the northern side by a factor of $\sim 2$. A similar asymmetry in the distribution of the scattered light has been extensively studied for the HH 30 disk, i.e., a low-mass protostar at a similar evolutionary stage, using the {\it Hubble Space Telescope} \citep[e.g.,][]{Burrows96,Stapelfeldt99,Cotera01,Watson07_HH30}. These authors attributed such a distribution to uneven illumination by the central object. These observations also show that the distribution of the scattered light is time variable.  \citet{Watson07_HH30} summarize possible mechanisms to explain the asymmetric distribution and its time variation. These include (1) hot (or cool) spots on the star; (2) shadowing by a non-axisymmetric inner disk; (3) obscuration by a companion star; and (4) obscuration by a disk associated with a companion star. The time variability in the scattered light is also observed in the more evolved disk system HD 163296 \citep{Wisniewski08}. The authors attributed this to the asymmetric shadowing of the inner disk.

Of the explanations described above, shadowing by the inner disk may be the most likely for the asymmetric $PI$ distribution in RY Tau. While uneven illumination by hot (or cool) spots is one of the favored explanations for the HH 30 disk \citep{Wood98,Stapelfeldt99,Cotera01,Watson07_HH30}, it is not likely for RY Tau: the optical continuum associated with this star does not show clear evidence for time variability due to such spots \citep[e.g.,][]{Petrov99,Chou13}. Obscuration by a secondary star/disk may also be excluded, since 
there is no clear evidence for the presence of a close companion associated with RY Tau. Near-infrared interferometric observations have ruled out the presence of a companion at 0.35--4 AU from the star and down to a stellar flux ratio of 0.05 \citep{Pott10}.

The shadowed-disk scenario is corroborated by the UXOR-type variability observed toward this star, i.e., time-variable obscuration by circumstellar dust \citep[e.g.,][]{Herbst94,Petrov99}. While many authors attribute this obscuration to the disk, its nature is not clear. It could be clumpy accretion of dust+gas onto the star \citep[see][and references therein]{Herbst94,Grady00}, the puffed-up inner rim of the disk \citep{Dullemond03}, or the outer edge of the disk \citep{The94}. \citet{Herbst94} also state the possibility that such a dust screen is associated with a wind, not a disk or disk accretion. Of the above explanations, the outer edge of the disk may be excluded if the disk associated with RY Tau is geometrically thin, and we are not observing this target close to an edge-on view (Sections 4 and 6.1).

RY Tau showed two abrupt brightening events at optical wavelengths in 1983/1984 and 1996/1997 \citep[e.g.,][]{Herbst94,Petrov99}. If we attribute this to a single orbital period around a 2 $M_\sun$ star, the corresponding radius of the disk is 7 AU. This radius is comparable to the radius where \citet{Isella10} identified emission peaks at two sides of the disk using millimeter interferometry. If obscuration occurs in the inner disk, it would require time variation of the disk structures including warping and precession. This may be possible via tidal interaction with a very low-mass companion such as a proto-planet \citep{Hughes09}.



Synoptic observations of the scattered light of RY Tau would allow us to identify the origin of the asymmetric $PI$ distribution, determine a typical radius where the obscuration occurs, and constrain the disk geometry (and its time variation) within the coronagraphic mask. \citet{Bastien82} made monitoring observations of optical polarization integrated over the object, revealing variation of the position angle of polarization between --25$^\circ$ and 45$^\circ$ over a few years\footnote{The aperture size used for this study ranges between 8 and 20 arcsec, and this fact may also cause different polarization angles.}.
Spatially-resolved observations like ours would have significant advantages for searching for a periodic variability toward this active and complicated object. We also note that the coronagraphic observations for the optical $I$ image of RY Tau were made in 2007--2009, and these show a brighter lobe in the north than the south-west (McCleary et al., unpublished) in the outer envelope, i.e., the opposite side as we observed in the $PI$ distribution. Although these observations show scattered light in significantly more distant outer regions, this is consistent with the idea of time variation of scattering light associated with this object.



\section{Conclusions}

We present near-infrared coronagraphic imaging polarimetry of RY Tau. The scattered light in the circumstellar environment was imaged at $H$-band with a high resolution ($\sim$0".05) using Subaru-HiCIAO. 
The observed $PI$ distribution shows an angular scale of $\sim$1".0 ($\sim$140 AU) and $\sim$0".6 ($\sim$80 AU) along the major and minor axes of the disk, respectively, exhibiting a butterfly-like shape. The angular scale along the major axis is similar to that of the disk measured using millimeter interferometry.
The bright part of the emission is offset from the star toward the direction of the blueshifted jet. 
Such a distribution can be explained if the object is associated with a geometrically thick disk or a remnant envelope. This agrees with the premise that the system is at the earliest stage of the Class II evolutionary phase.

We perform comparisons between the observed $PI$ distribution and simulations of scattered light with conventional disk and dust models. The simulated images are made using two different approaches. The first approach is to conduct full radiative transfer simulations including absorption, scattering and re-radiation by dust grains, with disk parameters based on SED fitting \citep[][]{Robitaille06,Robitaille07}. The second is monochromatic simulations with absorption and scattering only for a significantly larger sets of disk parameters to attempt to better fit the observed $PI$ distribution. 

The first approach reproduces the $PI$ flux level normalized to the integrated stellar $I$ flux well, but fails to reproduce the offset of the bright $PI$ distribution from the star along the disk axis. The second approach reproduces the butterfly-like morphology in $PI$ distribution well, with a total optical thickness of 1 at 30$^\circ$ from the midplane of the disk and the viewing angle of $\sim 50 ^\circ$. However, the model $PI$ distribution does not match the observations with a large viewing angle inferred by millimeter interferometry ($> 65^\circ$).
%
These results for the second approach are relatively independent of the free parameters for the disks and the size distribution of dust grains.

Throughout, we find disagreements between the observed $PI$ distribution and models using conventional disk and dust models. This may imply that the system consists of the following: (1) a geometrically thin disk which is partially responsible for the infrared SED but not the $PI$ flux in the near infrared; and (2) an optically thin and geometrically thick upper layer which is responsible for the observed $PI$ distribution in near infrared and the remaining infrared flux. 
Simulations show that this idea approximately explains the observed $PI$ emission with a viewing angle consistent with the observations. Half of the scattered light and thermal radiation in this layer illuminates the disk surface, and this process may significantly affect the thermal structure of the disk.

The $PI$ brightness has an asymmetry about the jet axis by a factor of $\sim$2 in flux presumably due to uneven illumination caused by obscuration by the dusty environment. Such obscuration could be either due to the accretion of dust+gas onto the star, the puffed-up inner rim of the disk, or the outer edge of the disk. Synoptic studies of the $PI$ distribution would give useful constraints for the geometry of the disk and perhaps its time variation within 5 AU of the star, and a better understanding of the nature of objects with a similar variability at optical wavelengths (UXORs).

\acknowledgments
We are grateful for an anonymous referee for a thorough review and valuable comments.
We thank support from the Subaru Telescope staff, especially from Michael Lemmen for making our observations successful.
We thank Drs. Shigehisa Takakuwa and Yasuhiro Hasegawa  for useful discussion. This research made use of the Simbad data base operated at CDS, Strasbourg, France, and the NASA's Astrophysics Data System Abstract Service. MT is supported from National Science Council of Taiwan (Grant No. {100-2112-M-001-007-MY3}). JPW is supported by NSF-AST 1009314. JC was supported by NSF-AST 1009203. CAG acknowledges support under NSF AST 1008440.



{\it Facilities:} \facility{Subaru (HiCIAO)}.



\appendix


\section{Simulated results with different carbon dust models}

Table \ref{tbl_dust_properties_appendix} shows the dust properties of the C01 and C01$\times$15 size distributions (Section 5.1) with different carbon dusts: i.e., amorphous carbon with pyrolysis temperature 400--1000 $^\circ$C and graphite. The optical constants for the amorphous carbon and graphite are taken from \citet{Jager98} and \citet{Draine84}, respectively. Figure \ref{scat_properties_appendix} shows their scattering properties as in Figure \ref{scat_properties}, but using the above carbon dusts.
%
The simulated $PI$ images using these dust grains show nearly identical morphologies, with differences in $PI$ flux of --20 to +30 \% compared with the amorphous carbon used in the main text (pyrolysis temperature = 600 $^\circ$C). As discussed in Section 5.2, these similarities and differences are attributed to ($PI/I_0$) at scattering angles 60--135$^\circ$ in Figure \ref{scat_properties_appendix}.

\section{Polarization, disk masses and extinction toward the star for geometrically thick disk models}

Figure \ref{Sprout_different_size_distributions_polvec} shows polarization vectors for the same modeled parameters as Figure \ref{Sprout_different_size_distributions}. For all the models the vector pattern is centrosymmetric about the star as it is in the observations (Section 3). We measure a degree of the polarization of 10--28 \% at the position corresponding to the $PI$ peak of the observations (see Figure  \ref{Sprout_different_size_distributions} for the position). Some seem inconsistent with the observations ($\gg$9 \%, Section 3).

Tables \ref{disk_masses} and \ref{A_V} show the disk masses and extinction ($A_V$) toward the star, respectively, for the models used in Section 5.2. The disk masses are derived assuming a gas-to-dust mass ratio of 100. The extinctions are calculated based on the optical thickness at 1.65 $\micron$ and the $\kappa_{\rm ext; 1.65 \micron} / \kappa_{\rm ext; 0.55 \micron}$ tabulated in Table \ref{tbl_dust_properties}.
Table \ref{disk_masses} show that the disk mass is significantly smaller for $h_{50AU}=25$ AU than $h_{50AU}=5$ AU with a given combination of the dust model, $\beta$, and $\theta_{\tau=1}$. As a result, the former disks produce brighter $PI$ flux from the other side of the disk (Figures \ref{Sprout_30deg},  \ref{Sprout_25deg}, \ref{Sprout_different_size_distributions}). Disks with a large scale height also provide a larger extinction toward the star.

All the disk masses derived for $h_{50AU}=25$ AU are significantly smaller than that estimated by \citet{Isella10} based on interferometry ($>3 \times 10^{-3} M_\sun$). In contrast, a majority of the disk masses derived for $h_{50AU}=5$ AU are significantly larger than those of pre-main sequence stars suggested by millimeter observations \citep[$< 0.1 M_\sun$, e.g.,][see also Table 1]{Robitaille07,Isella09,Williams11}. However, we do not constrain the models described in Section 5.2 for the following reasons: (1) the disk mass inferred by millimeter interferometry highly depends on the dust model used to convert the flux to a dust mass \citep{Isella10}; (2) the modeled disk mass highly depends on the exponent of the radial density distribution $\alpha$, which is assumed to be $\beta+1$ for the simulations in Section 5.2; and (3) the scattered light results from a small fraction to the total dust mass (Appendix C), therefore our observations do not directly probe the total disk mass.

Regarding the extinction to the star, some models provide a larger value than that measured by \citet{Calvet04} ($A_V$=2.2). However, we note that the measurement of extinction by \citet{Calvet04} is based on observations at UV-optical wavelengths, adopting an extinction law for cold molecular clouds. Therefore, it may not be directly comparable with the extinctions of Table \ref{A_V}, in particular for the C01 and C01$\times$15 size distributions.
 

\section{Optical thickness, density and mass of the optically thin and geometrically thick scattering layer}
Suppose the dust corresponding to the mass $m$ is located at distance $r$ from the star. The number of $PI$ photons observed at the telescope per second is described as follows:
\begin{equation}
n_{PI} = \kappa_{ext} m  \frac{N_*}{4 \pi r^2} \left( \frac{PI}{I_0} \right) \Omega,
\end{equation}
where $\kappa_{ext}$ is the opacity; $N_*$ is number of the photons at 1.65 $\mu$m ejected from the star in all directions; and ($PI/I_0$) is the fraction of the $PI$ flux normalized to the incident flux on dust grains (Figure 4). $\Omega$ is the solid angle corresponding to the telescope area, thus $\Omega = A_{tel}/d^2$, where $A_{tel}$ and $d$ are the area of the telescope and the distance to the target, respectively. If we normalize Equation (2) by the number of stellar photons observed per second, i.e., $n_* = N_* (A_{tel}/4 \pi d^2$),  the equation is as follows:-
\begin{equation}
\frac{n_{PI}}{n_*} = \frac{\kappa_{ext} m}{r^2} \left( \frac{PI}{I_0} \right),
\end{equation}

\begin{figure*}
\epsscale{2.2}
\plotone{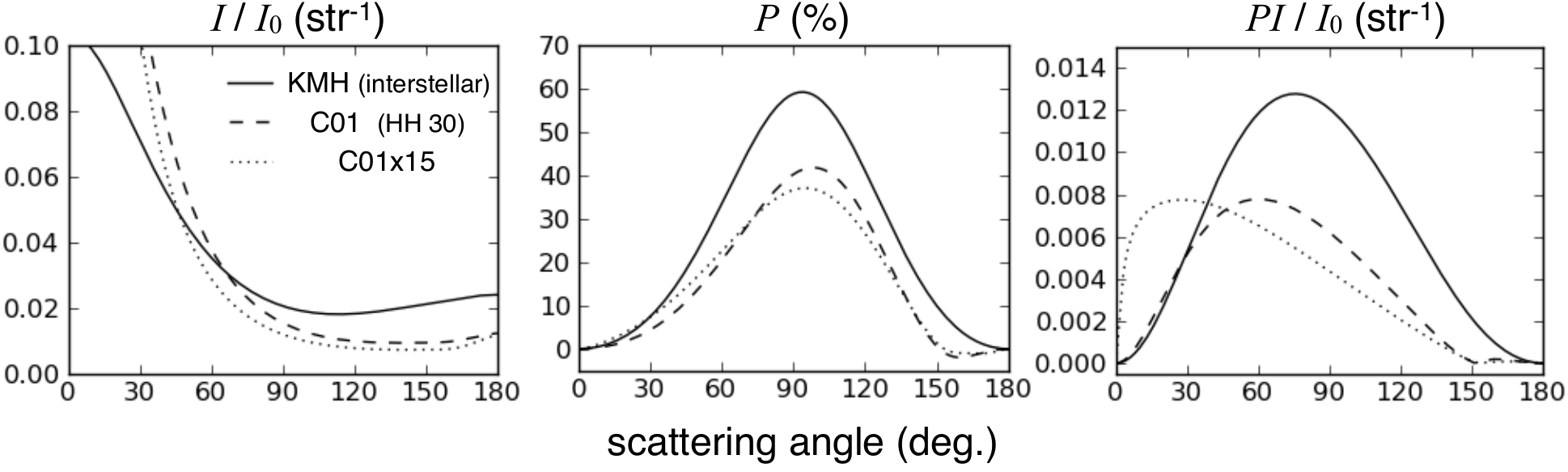}
\caption{Scattering properties of an unpolarized photon for different size distributions. From left to right the figures show the scattered $I$ flux per steradian, the polarization, and the scattered $PI$ flux per steradian, as a function of scattering angle. The $I$ and $PI$ fluxes are normalized to the incident Stokes $I$ parameter ($I_0$), and as a result, $\int I/I_0 ~d\Omega =$ scattering albedo.
\label{scat_properties}}
\end{figure*}

\begin{figure}
\epsscale{1.1}
\plotone{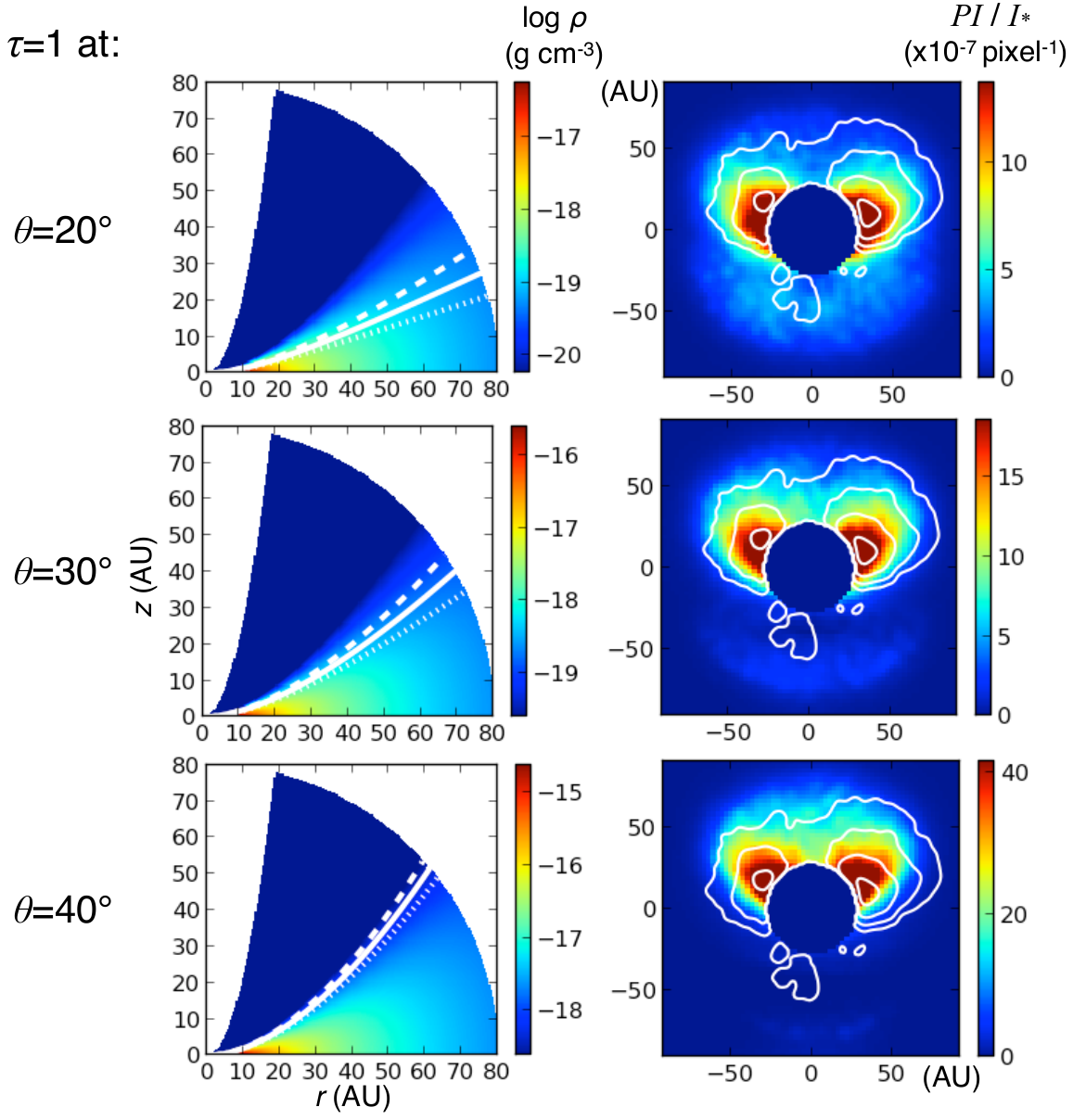}
\caption{Density distribution of dust (left) and monochromatic simulations of the $PI$ distribution (right) for $\tau = 1$ at $\theta = 20/30/40 ^\circ$ (top/middle/bottom),  $h_{\rm 50 AU}$=15 AU, $\beta$=2.0, viewing angle of 49$^\circ$ from the pole, and the C01 dust size distribution.
The color scale is arbitrarily adjusted to clearly show distributions similar to the contours, which show the observed $PI$ distribution. The contour scales are 0.075, 0.15, 0.3 and 0.6$\times 10^{-6}$ per pixel scale of HiCIAO relative to the stellar flux, and they are rotated by 72$^\circ$ from the observations. Dashed, solid and dotted curves show the positions for $\tau$=0.5, 1, and 2 from the star, respectively.
\label{Sprout_examples}}
\end{figure}

\begin{figure*}
\epsscale{2.2}
\plotone{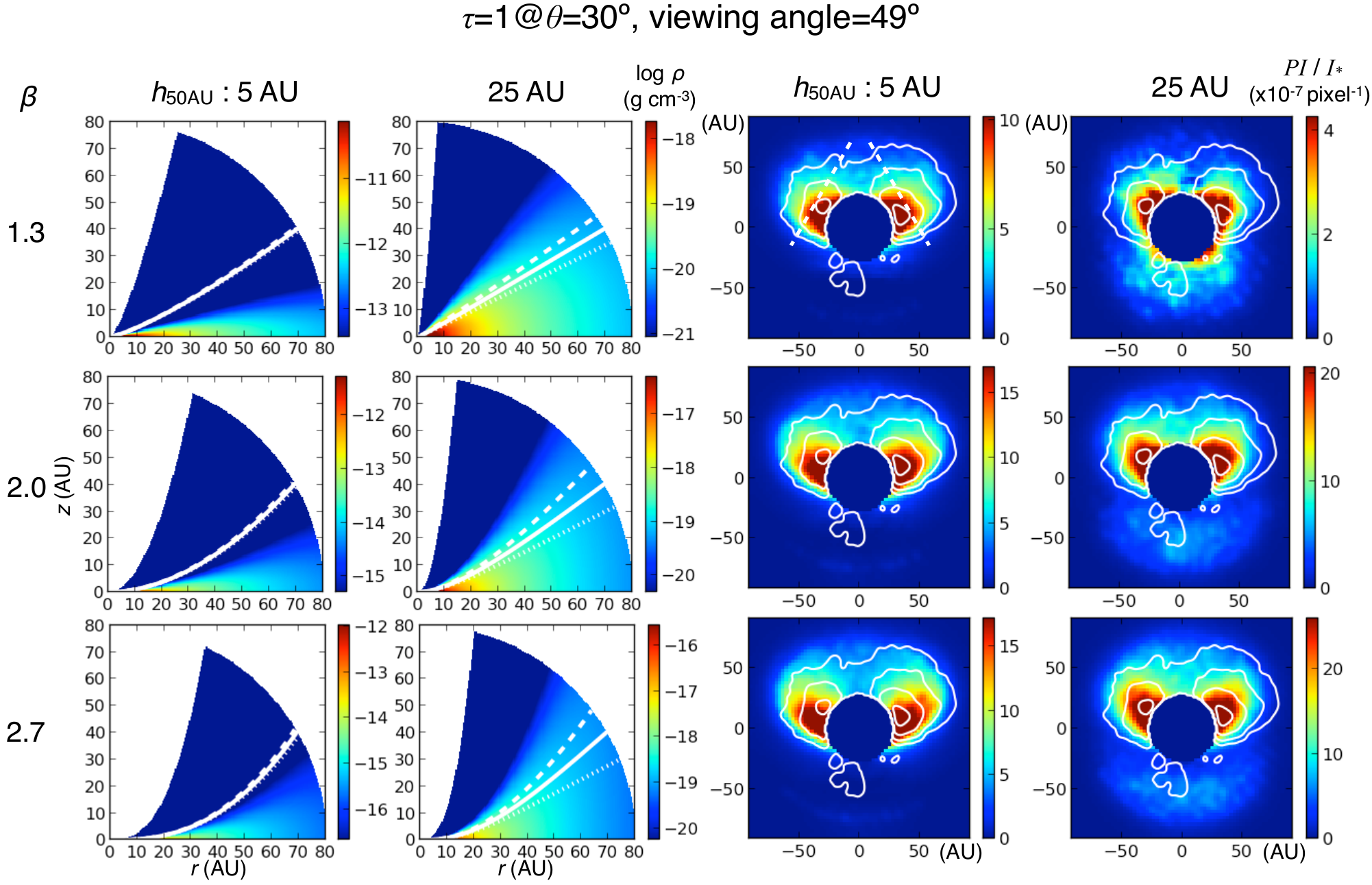}
\caption{Same as Figure \ref{Sprout_examples} but for $\tau = 1$ at $\theta = 30 ^\circ$ with different $h_{\rm 50 AU}$ and $\beta$. White dashed lines in the top-right $PI$ image show the positions where we extract the 1-D distribution shown in Figure \ref{Sprout_profs}.
\label{Sprout_30deg}}
\end{figure*}

\begin{figure*}
\epsscale{2.2}
\plotone{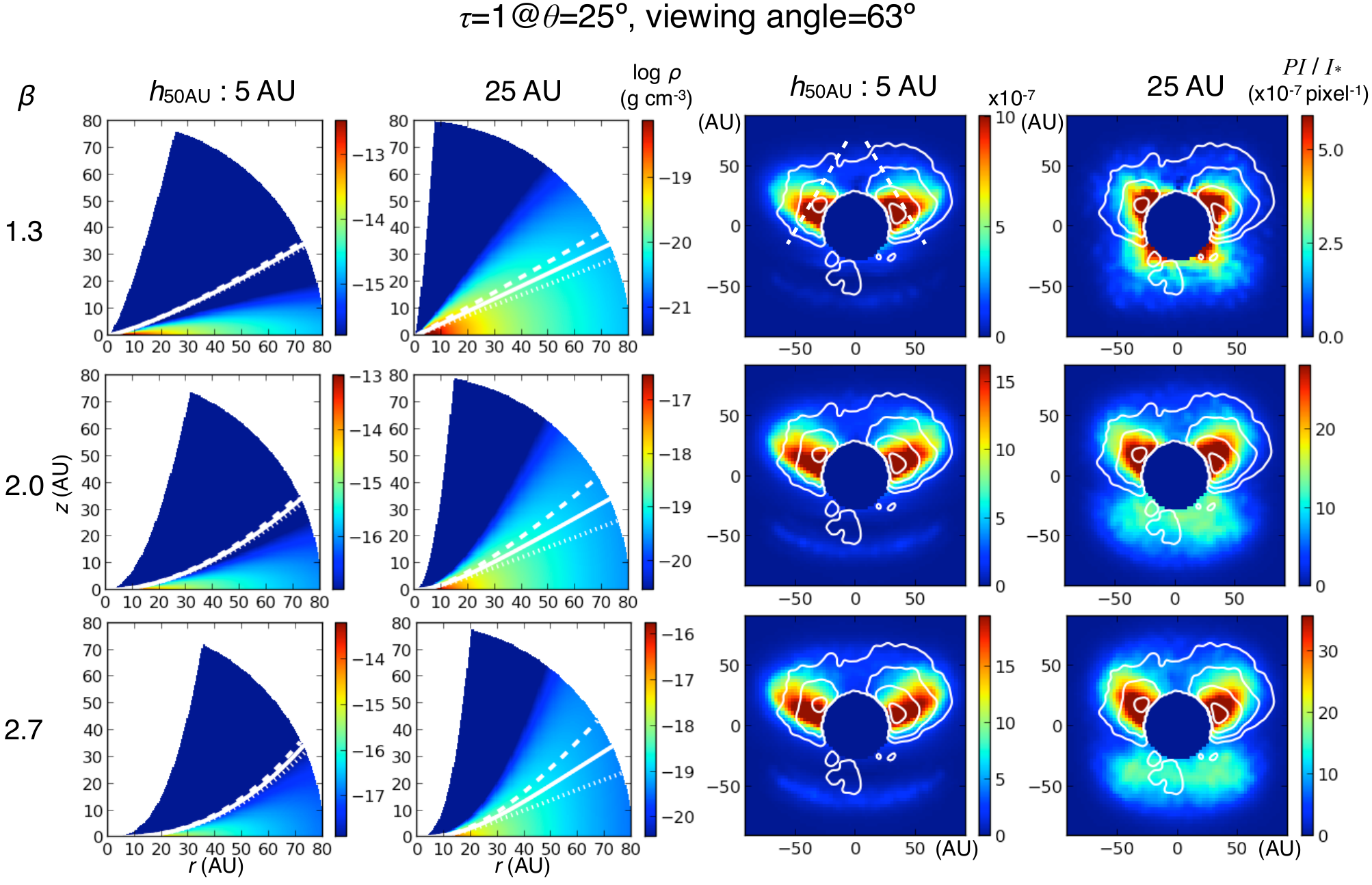}
\caption{Same as Figure \ref{Sprout_30deg} but for $\tau = 1$ at $\theta = 25 ^\circ$ and a viewing angle of 63$^\circ$ from the face-on view.
\label{Sprout_25deg}}
\end{figure*}

\begin{figure}
\epsscale{1.0}
\plotone{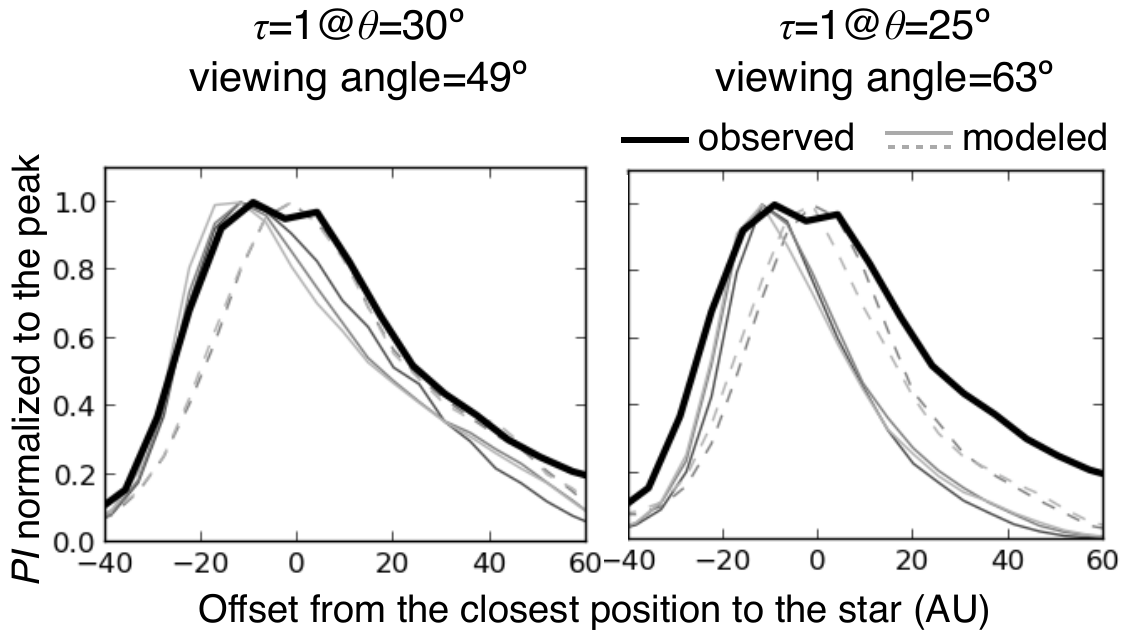}
\caption{The 1-D $PI$ distribution extracted from the observed (black) and modeled $PI$ images (gray). The positions of the extraction are shown in Figures \ref{Sprout_30deg} and \ref{Sprout_25deg}. Solid and dashed gray curves are models for $h_{\rm 50 AU}$=5 and 25 AU, respectively. Each profile is made by averaging those at the two sides and normalizing to the peak $PI$ flux.
\label{Sprout_profs}}
\end{figure}

\begin{figure*}
\epsscale{2.2}
\plotone{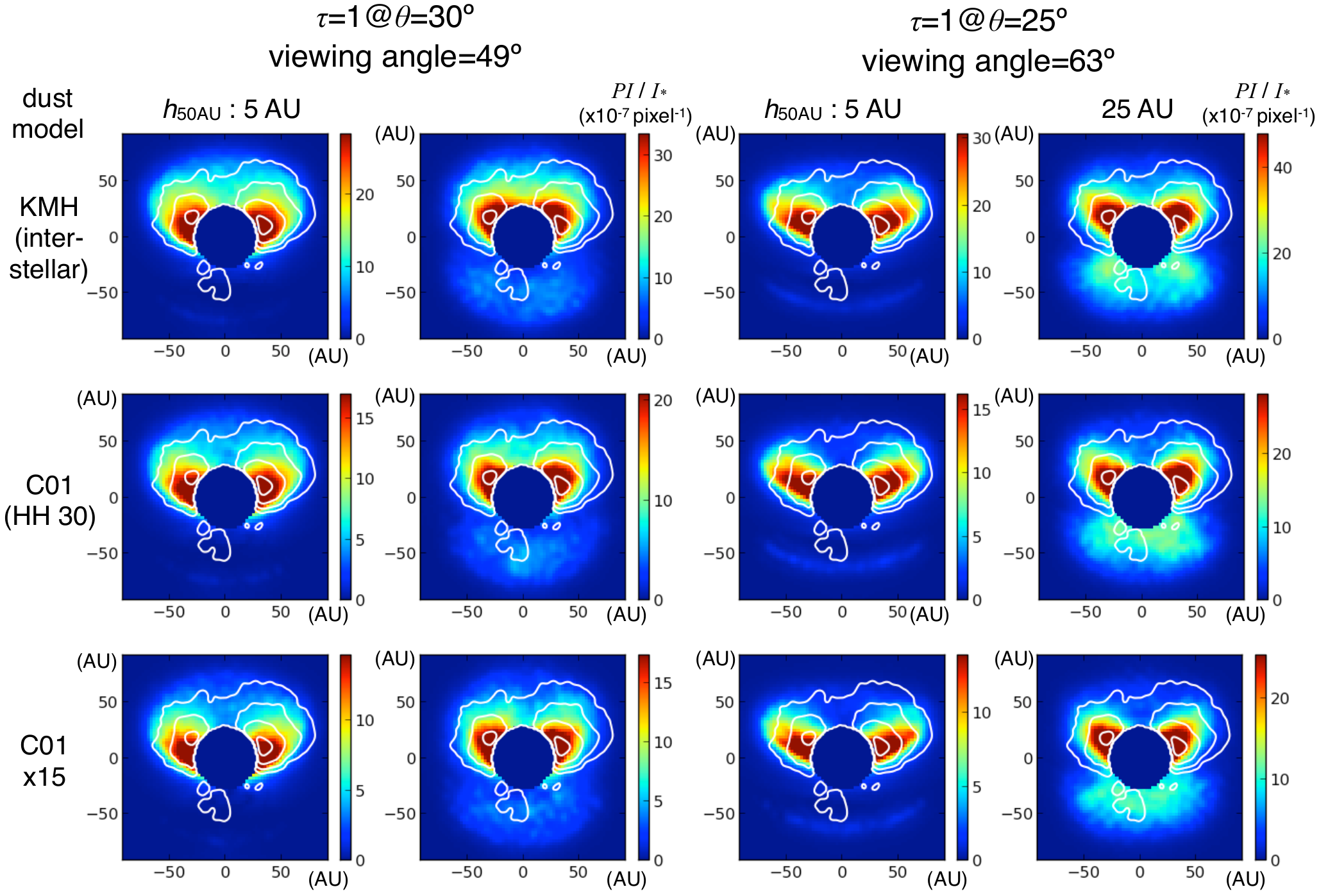}
\caption{
$PI$ distributions with different dust models for $\beta$=2.0. The other parameters ($\theta$, viewing angle, and $h_{\rm 50 AU}$) and contours are the same as Figures \ref{Sprout_30deg} and \ref{Sprout_25deg}. 
\label{Sprout_different_size_distributions}}
\end{figure*}

\begin{figure}
\epsscale{1.0}
\plotone{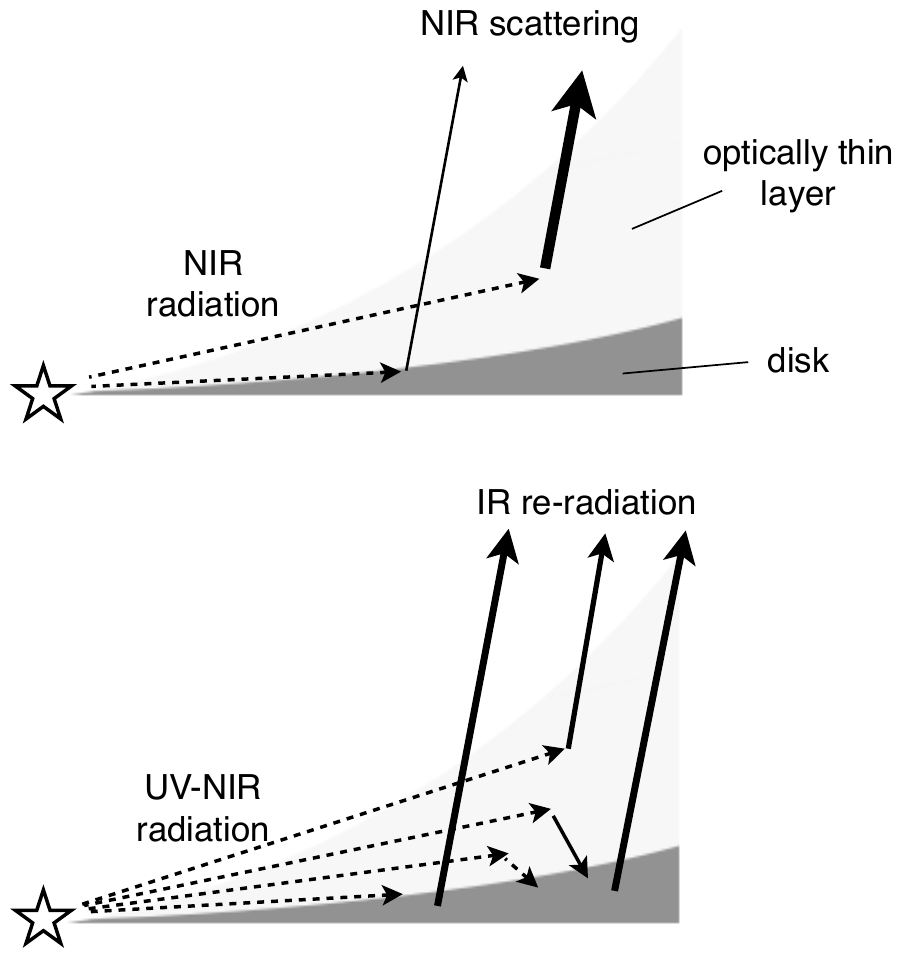}
\caption{A schematic view of the disk, optically thin layer, stellar radiation, and the near-infrared (NIR) scattered light and infrared re-radiation observed. See text for details.
\label{disk_sketches}}
\end{figure}

\begin{figure}
\epsscale{1.1}
\plotone{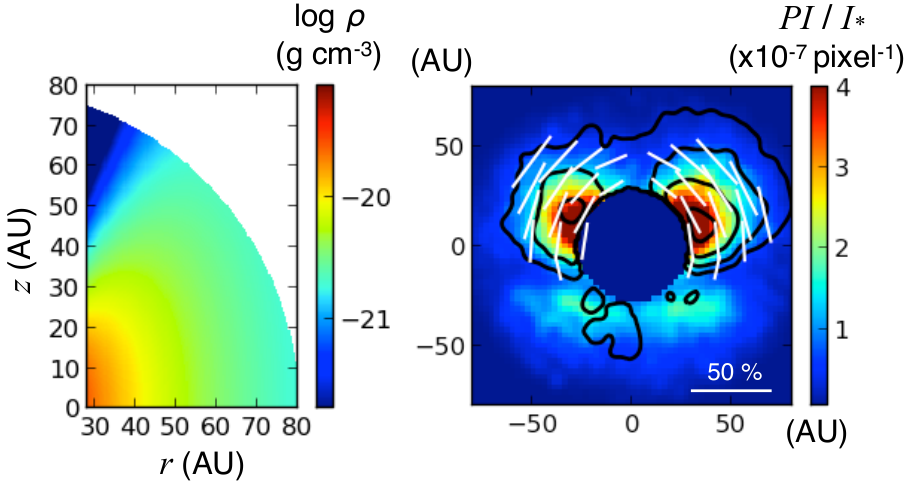}
\caption{A model with a disk with zero-thickness and an optically thin and geometrically thick upper layer using the KMH dust model. ($left$) Density distribution of the optically thin layer ($r$=28--80 AU). This is set using Equation (1) with $\alpha$=2.5, $\beta$=1.5, $h_{50AU}$=30 AU, and a total optical thickness of $\tau = 0.04$ at 1$^\circ$ above the midplane. The total dust mass of the layer is $7.3 \times 10^3~M_\earth$. ($right$) Modeled $PI$ distribution (color) and polarization vectors (lines) with a viewing angle of 70$^\circ$. $3 \times 10^6$ photons were used for the simulation. The color contrast for the $PI$ distribution is same contrast as Figure \ref{fig1}. The black contours show the observed $PI$ distribution.  
\label{Sprout_opt_thin}}
\end{figure}

\begin{figure*}
\epsscale{2.2}
\plotone{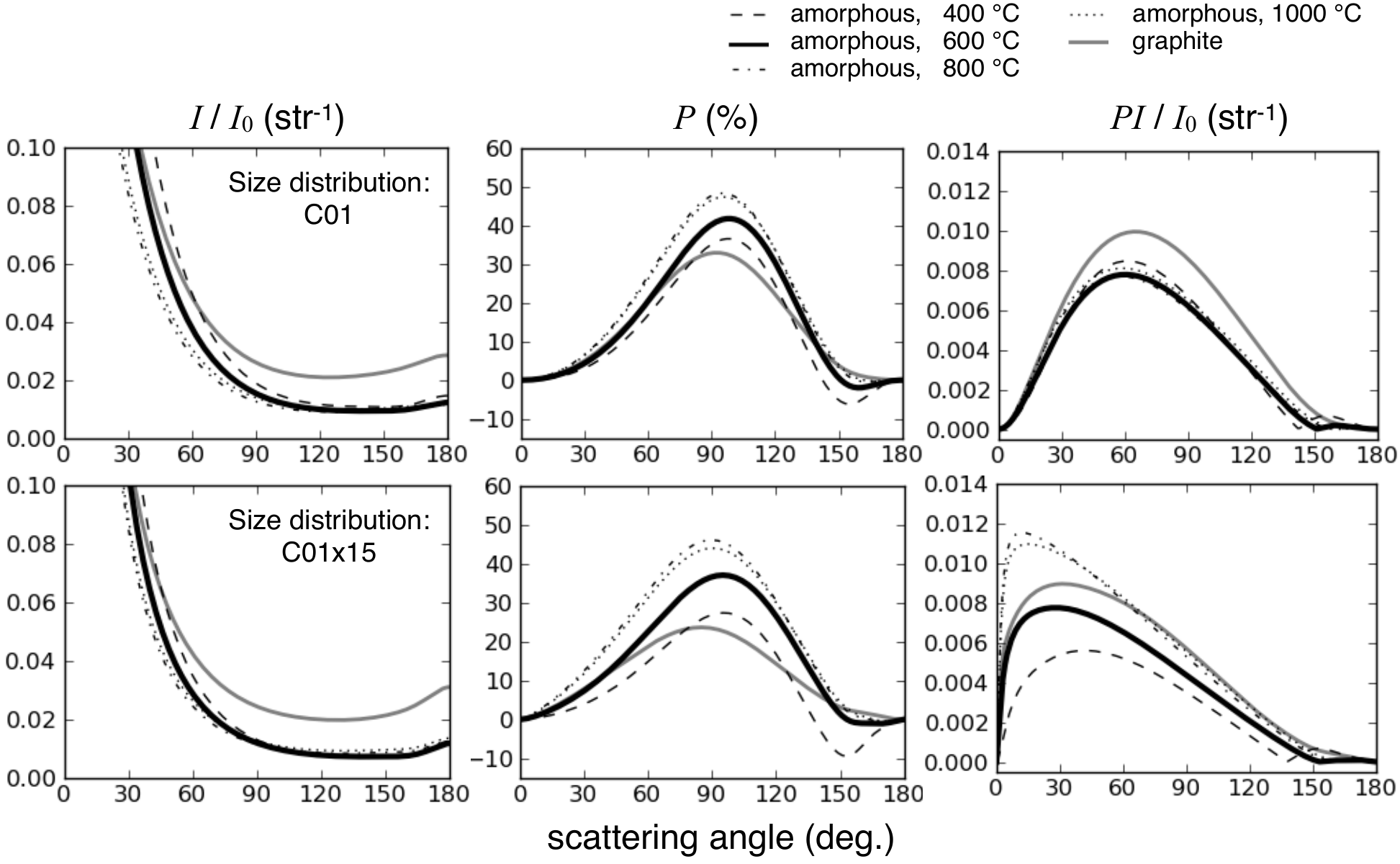}
\caption{Same as Figure \ref{scat_properties} but with the C01 (upper) and C01$\times$15 size distributions (lower) and different carbon dusts (amorphous carbon with pyrolysis temperatures 400/600/800/1000 $^\circ$C and graphite). The black and gray curves are for amorphous carbon and graphite, respectively. Those for amorphous carbon with different pyrolysis temperatures are shown with different line styles, and that used in Section 5 (pyrolysis temperature of 600 $^\circ$C) is shown with black thick curves. 
\label{scat_properties_appendix}}
\end{figure*}

\begin{figure*}
\epsscale{2.2}
\plotone{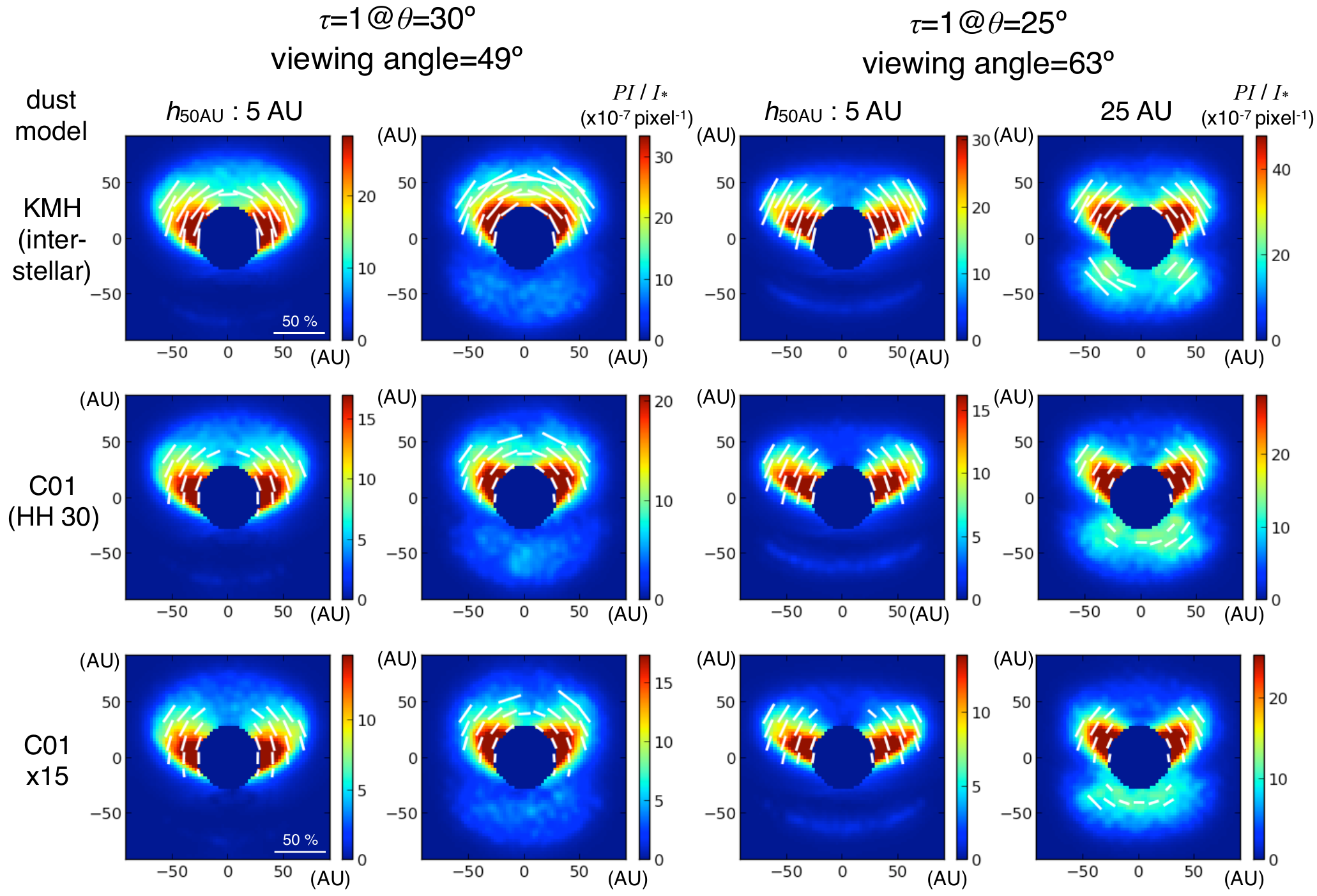}
\caption{
Same as Figure \ref{Sprout_different_size_distributions_polvec} but with the polarization vectors and without the contour for the observed $PI$ distribution.
\label{Sprout_different_size_distributions_polvec}}
\end{figure*}

Here we substitute $n_{PI}/n_* = 8\times10^{-7}$, measured at the peak of the $PI$ flux as described in Section 3; $\kappa_{ext}$ = $1.1 \times 10^4$ cm$^2$ g$^{-1}$ (C01 dust, Table \ref{tbl_dust_properties}), $r$=40 AU, and $PI/I_0 = 5\times10^{-3}$ str$^{-1}$ (Figure \ref{scat_properties}). We derive a dust mass of $5\times10^{21}$ g. The $PI$ flux measured at each position is based on the HiCIAO pixel scale ($9.48 \times 10^{-3}$ arcsec), thus this dust mass corresponds to an optical thickness along the line of sight of $\tau = \kappa_{ext} m / A_{pix} \sim 0.15$, where $A_{pix}$ is the area corresponding to each pixel, assuming a distance $d$=140 pc. Therefore, it is likely that the scattering layer is optically thin along the line of sight.

Such a layer could also be optically thin in the radial direction. The optical thickness along this direction highly depends on the assumed geometrical thickness of the layer. Let us assume a thickness of 30 AU, comparable to the thickness of the disks we used in Section 5 to explain the offset of the $PI$ distribution from the star.
One would expect an optical thickness of $\sim 6 \times 10^{-3}$ towards a HiCIAO pixel (1.3 AU) assuming all of the parameters described above. Integrating this over the disk radius of 80 AU, one would expect a total optical thickness of $\sim 0.4$. This is a simple estimate. In reality, the observed $PI$ flux is lower than the maximum at the other positions, and the dependence of $n_{PI}/n_*$ on the distance from the star should be included. Furthermore, the opacity at individual positions depends on the distance from the star (Equation 3), and this fact is not included here.

If we assume the same layer thickness, and use the dust mass contained in the pixel with the maximum $PI$ value, we would derive a dust mass density of $2.9 \times 10^{-20}$ cm$^{-3}$. This corresponds to a hydrogen number density of the order of $\sim 10^6$ cm$^{-3}$ assuming a gas-to-dust mass ratio of 100. This is larger than the envelope density at similar radii ($\sim$50 AU) inferred by the SED fitting in Section 4 by a factor of $\sim$10 or more.

We estimate the dust mass of the scattering layer as follows, using Equation (3):-
\begin{equation}
M_{\rm dust} = \int m({\bf r}) d{\bf r} = \int \frac{n_{PI} ({\bf r})}{n_*} \frac{r^2}{\kappa_{ext}} \left( \frac{PI}{I_0} \right)^{-1} d{\bf r}
\end{equation}
where {\bf r} and $r$ are the position and the distance to the star, respectively. Substituting the projected radius for $r$, we derive a dust mass $M_{\rm dust}$ of $3-7 \times 10^{-3}~M_\earth$.  A larger value would be expected if we include the inclination effect for $r$ and the smaller $PI/I_0$, depending on the scattering angle at individual positions. 
Even so, this estimate would be sufficient to conclude that the optically thin layer above the disk surface has a mass significantly smaller than the total dust mass in the disk inferred by radio observations \citep[$10-50~M_\earth$,][]{Isella10}.

\clearpage



\clearpage

\begin{table*}
\begin{center}
\caption{UV-to-Radio Fluxes\tablenotemark{a}\label{tbl_observed_SED}}
\begin{tabular}{cccc}
\tableline\tableline
Wavelength & Flux  & Year of           &Reference\tablenotemark{e} \\
($\micron$)   & (mJy) & Observations &                  \\
\tableline
0.36	&	43 $\pm$ 19\tablenotemark{b}     			& 1962-1990\tablenotemark{b}		& 1\\
0.44	&	$(1.5 \pm 0.6) \times 10^2$\tablenotemark{b} 	& 1962-1990\tablenotemark{b}		& 1\\
0.55	&	$(3.5 \pm 1.5) \times 10^2$\tablenotemark{b} 	& 1962-1990\tablenotemark{b}		& 1\\	
0.64	&	$(5.1 \pm 2.2) \times 10^2$\tablenotemark{b} 	& 1962-1990\tablenotemark{b}		& 1\\	
0.71	&	$(8.9 \pm 3.9) \times 10^2$\tablenotemark{b} 	& 1962-1990\tablenotemark{b}		& 1\\	
1.25	&	$(2.26 \pm 0.07) \times 10^3$ 				& 1997--2001\tablenotemark{c}	& 2\\	
1.65	&	$(3.8 \pm 0.2) \times 10^3$ 				& 1997--2001\tablenotemark{c}	& 2\\	
2.2	&	$(4.7 \pm 0.1) \times 10^3$ 				& 1997--2001\tablenotemark{c}	& 2\\	
3.6	&	$(5.75 \pm 0.07) \times 10^3$ 				& 2003--2007\tablenotemark{d}	& 3\\	
4.5	&	$(5.30 \pm 0.06) \times 10^3$ 				& 2003--2007\tablenotemark{d}	& 3\\	
5.8	&	$(4.4 \pm 0.2) \times 10^3$ 				& 2003--2007\tablenotemark{d}	& 3\\	
8.0	&	$(5.8 \pm 0.2) \times 10^3$ 				& 2003--2007\tablenotemark{d}	& 3\\	
12	&	$(1.774 \pm 0.003) \times 10^4$ 			& 1983						& 4\\	
24	&	$(1.8 \pm 0.4) \times 10^4$ 				& 2003--2007\tablenotemark{d}	& 3\\	
25	&	$(2.648 \pm 0.005) \times 10^4$ 			& 1983						& 4\\	
60	&	$(1.891 \pm 0.007) \times 10^4$ 			& 1983						& 4\\	
70	&	$(9.6 \pm 1.0) \times 10^3$ 				& 2003--2007\tablenotemark{d}	& 3\\	
100	&	$(1.4 \pm 0.2) \times 10^4$ 				& 1983		& 4\\	
350	&	$(2.4 \pm 0.3) \times 10^3$ 				& 1990		& 5\\	
450	&	$(1.9 \pm 0.2) \times 10^3$ 				& 1990		& 5\\	
800	&	$(5.6 \pm 0.3) \times 10^2$ 				& 1990		& 5\\	
\tableline
\end{tabular}
\tablenotetext{a}{ Cited from \citet{Robitaille07}. The $L$- (3.5 $\micron$), $M$- (4.8 $\micron$), and $N$-band (10.5 $\micron$) data quoted from \citet{Kenyon95} are not included, as these have large uncertainties, and are consistent with other observations at similar wavelengths.}
\tablenotetext{b}{ Measurements at multiple epochs. The uncertainty is based on the photometric variability.}
\tablenotetext{c}{ The specific year is not clear. We describe the years of operation for the 2MASS all-sky survey.}
\tablenotetext{d}{ The specific year is not clear. We describe the period between the launch of the Spitzer Space Telescope and when \citet{Robitaille07} was published.}
\tablenotetext{e}{1 --- \citet{Herbst94}, and converted to the tabulated wavelengths by \citet{Robitaille07} ; 2 --- 2MASS all-sky survey; 3 --- Spitzer Space Telescope Archive; 4 --- \citet{Weaver92}; 5 --- \citet{Andrews05}, who quoted the values from \citet{Mannings94}}
\end{center}
\end{table*}



\begin{deluxetable}{cccccccccccccccc}
\tabletypesize{\scriptsize}
\rotate
\tablecaption{Best-Fit Models using the SED Fitter \label{best_fit_models}}
\tablewidth{0pt}
\tablehead{

\colhead{Model ID} &
\colhead{Inclination} &
\colhead{$\chi^2$} &
\colhead{}&
\multicolumn{2}{c}{Stellar Parameters} &
\colhead{}&
\multicolumn{6}{c}{Disk Parameters} &
\colhead{}&
\multicolumn{2}{c}{Others} \\


\cline{5-6} \cline{8-13} \cline{15-16} \\

\colhead{}&
\colhead{}&
\colhead{}&
\colhead{}&

\colhead{$R_*$} &
\colhead{$T_*$\tablenotemark{a}} & 
\colhead{}&

\colhead{$R_{\rm min}$} &
\colhead{$R_{\rm max}$} & 
\colhead{$h_{\rm 50AU}$} & 
\colhead{$\beta$} & 
\colhead{$M_{\rm disk}$} & 
\colhead{$\dot{M}_{\rm disk}$} & 
\colhead{}&

\colhead{$M_{\rm envelope}$} &
\colhead{$\rho_{\rm cavity}$} \\


\colhead{}&
\colhead{($^\circ$)}&
\colhead{}&
\colhead{}&

\colhead{($R_\sun$)} &
\colhead{(K)} & 
\colhead{}&

\colhead{($R_{\rm sub}$)} &
\colhead{(AU)} & 
\colhead{} & 
\colhead{} & 
\colhead{($M_\sun$)} & 
\colhead{($M_\sun$ yr$^{-1}$)} & 
\colhead{}&

\colhead{($M_\sun$)} &
\colhead{(g cm$^{-3}$)} 
}

\startdata

3000949

& 75.5 & 1.3$\times 10^2$ & &

4.9 & $4.7 \times 10^3$ & & 
1.0 & 83 & 2.4 & 1.13 & 1.1$\times 10^{-1}$  & 1.2$\times 10^{-6}$ & & 
1.1$\times 10^{-4}$  & 1.7$\times 10^{-21}$\\

 & 69.5 & 1.6$\times 10^2$ \\
 & 63.2 & 2.2$\times 10^2$\vspace{0.2cm}\\ 




3012376

& 69.5 & 1.6$\times 10^2$ & &

5.3 & $5.1 \times 10^3$ & & 
7.8 & 84 & 2.7 & 1.16 & 8.0$\times 10^{-2}$  & 6.1$\times 10^{-8}$ & & 
3.4$\times 10^{-4}$  & 2.0$\times 10^{-21}$\\

 & 63.3 & 1.8$\times 10^2$ \\
 & 75.5 & 2.0$\times 10^2$ \\
 & 56.6 & 2.3$\times 10^2$ \\ 

3007615

& 69.5 & 2.1$\times 10^2$& &

4.6 & $6.4 \times 10^3$ & & 
7.0 & $4.0 \times10^2$ & 1.6 & 1.12 & 4.6$\times 10^{-2}$  & 1.1$\times 10^{-7}$ & & 
2.0$\times 10^{-6}$  & 0 \\

 & 75.5 & 2.3$\times 10^2$\vspace{0.2cm}\\ 

3012808

& 56.6 & 2.3$\times 10^2$& &

5.0 & $5.1 \times 10^3$ & & 
4.7 & $1.2 \times 10^2$ & 2.1 & 1.17 & 2.8$\times 10^{-2}$  & 2.4$\times 10^{-7}$ & & 
3.5$\times 10^{-8}$  & 0\vspace{0.2cm}\\  

\enddata

\tablecomments{$R_*$ ... stellar radius;  $T_*$ ... stellar temperature; $R_{\rm min}$ ... minimum disk radius; $R_{\rm max}$ ... maximum disk radius; $M_{\rm disk}$ ... disk mass; $\dot{M}_{\rm disk}$ ... disk mass accretion rate; $M_{\rm envelope}$ ... envelope mass; $\rho_{\rm cavity}$ ... density in the outflow cavity; $R_{\rm sub}$ ... sublimation radius. 
The sublimation radius is empirically determined by $R_{sub} = R_* (T_{sub}/T_*)^{-2.1}$ \citep{Robitaille06}.
The inclination is measured from the polar axis. See text for definition of $\chi^2$. The masses and mass accretion rates are described for gas+dust assuming the gas-to-dust mass ratio of 100.}
\tablenotetext{a}{Any temperature listed here does not agree the measured spectral type of RY Tau of F8-G2 \citep{Petrov99,Calvet04,Mendigutia11a}, corresponding to an effective temperature of $5.8-6.3 \times 10^3$K. The different stellar temperatures listed here do not affect the modeled SEDs and $PI$ image at 1.65 $\micron$ (Section 4).}
\end{deluxetable}



\clearpage

\begin{table*}
\begin{center}
\caption{Physical Parameters for Dust Grains ($\lambda = 1.65 \micron$)\label{tbl_dust_properties}}
\footnotesize
\begin{tabular}{llcccccc}
\tableline\tableline
\multicolumn{2}{l}{Size distribution\tablenotemark{a}} & Mass fraction of & $\kappa_{\rm ext}$\tablenotemark{b}& $\kappa_{\rm ext} / \kappa_{\rm ext (0.55 \micron)}$ & albedo & $g$ & $P_{\rm max}$ \\
&& carbon dust & (cm$^2$ g$^{-1}$)\\
\tableline
KMH &(interstellar medium) 	&0.38	& $5.3 \times 10^3$ & 0.16	& 0.39 & 0.28 & 0.59	\\
C01 &(the HH 30 disk)          	&0.49	& $1.1 \times 10^4$ & 0.28	& 0.45 & 0.54 & 0.42	\\
C01$\times$15                    &	&0.49	& $5.4 \times 10^3$ & 0.50	& 0.47 & 0.66 & 0.37	\\
\tableline
\end{tabular}
\tablenotetext{a}{See text for details. The minimum and maximum particle radii are $3 \times 10^{-3}$ and 2 $\micron$ for KMH; $1 \times 10^{-3}$ and 20 $\micron$ for C01; and 15 times larger for C01$\times$15.}
\tablenotetext{b}{We adopt the mass density 3.3 and 2.26 g cm$^{-3}$ for silicate and graphite, respectively \citep{Kim94}, and 1.67 g cm$^{-3}$ for amorphous carbon \citep{Jager98}. This parameter is often defined per gas+dust mass \citep[e.g.,][]{Cotera01,Wood02b,Whitney03b,Whitney03a,Dong12a}, but in this paper we define this per dust mass to discuss the total dust mass in the scattering layer in Section 6.1. }
\end{center}
\end{table*}


\clearpage

\begin{deluxetable}{lrccccc} 
\tablecolumns{7} 
\tablecaption{Maximum $PI/I_*$\label{max_PI}} 
\tablewidth{0pt} 
\tablehead{
\colhead{Dust} &
\colhead{} &
\multicolumn{2}{c}{$\tau = 1$ at $\theta=30^\circ$} &
\colhead{} &
\multicolumn{2}{c}{$\tau = 1$ at $\theta=25^\circ$} \\
\cline{3-4} \cline{6-7} \\
\colhead{} &
\colhead{$\beta$} &
\colhead{$h_{50 \rm AU}$=5 AU} &
\colhead{25 AU} &
\colhead{} &
\colhead{5 AU} &
\colhead{25 AU}
}
\startdata 
KMH&1.3&
$3.0 \times 10^{-6}$& --- && $3.5 \times 10^{-6}$ &---\\
&2.0&
$5.1 \times 10^{-6}$ &$6.5 \times 10^{-6}$&&$5.0 \times 10^{-6}$&$9.4 \times 10^{-6}$\\
&2.7&
$4.4 \times 10^{-6}$ &$7.3 \times 10^{-6}$&&$4.5 \times 10^{-6}$&$9.3 \times 10^{-6}$
\vspace{0.1cm}\\
C01&1.3&
$1.9 \times 10^{-6}$&---&&$2.1 \times 10^{-6}$&---\\
&2.0&
$3.1 \times 10^{-6}$&$4.0 \times 10^{-6}$&&$2.7 \times 10^{-6}$&$5.7 \times 10^{-6}$\\
&2.7&
$2.6 \times 10^{-2}$&$4.1 \times 10^{-6}$&&$2.5 \times 10^{-6}$&$6.1 \times 10^{-6}$
\vspace{0.1cm}\\
C01$\times$15&1.3&
$1.8 \times 10^{-6}$& --- &&$1.8 \times 10^{-6}$& ---\\
&2.0&
$2.6 \times 10^{-6}$ &$3.3 \times 10^{-6}$&&$2.3 \times 10^{-6}$&$5.4 \times 10^{-6}$\\
&2.7&
$2.1 \times 10^{-6}$&$3.5 \times 10^{-6}$&&$2.0 \times 10^{-6}$&$5.5 \times 10^{-6}$
\enddata 
\end{deluxetable}


\begin{table*}
\begin{center}
\caption{Physical Parameters for Dust Grains with Different Carbon Dust ($\lambda = 1.65 \micron$)\label{tbl_dust_properties_appendix}}
\footnotesize
\begin{tabular}{llccccccc}
\tableline\tableline
Size distribution & Carbon & pyrolysis temperature  &	Mass fraction of		
  & $\kappa_{\rm ext}\tablenotemark{b}$& albedo & $g$ & $P_{\rm max}$ \\
  
                          &              & (amorphous only, $^\circ$C)	& carbon dust & (cm$^2$ g$^{-1}$) \\
\tableline
C01			& amorphous	& 400 	& 0.45	& $8.4 \times 10^3$	& 0.59	& 0.57	& 0.37	\\
			&			& 600	& 0.49	& $1.1 \times 10^4$	& 0.44	& 0.54	& 0.42	\\
			&			& 800	& 0.52	& $1.4 \times 10^4$	& 0.34	& 0.50	& 0.48	\\
			&			& 1000	& 0.53	& $1.4 \times 10^4$	& 0.36	& 0.49	& 0.47	\\
			& graphite 	& ---		& 0.57	& $1.1 \times 10^4$	& 0.57	& 0.39	& 0.33	\vspace{0.1cm} \\
C01$\times$15	& amorphous	& 400 	& 0.45	& $4.8 \times 10^3$	& 0.58	& 0.69	& 0.27 	\\
			&			& 600	& 0.49	& $5.4 \times 10^3$	& 0.47	& 0.66	& 0.37	\\
			&			& 800	& 0.52	& $6.2 \times 10^3$	& 0.41	& 0.61	& 0.46	\\
			&			& 1000	& 0.53	& $6.0 \times 10^3$	& 0.43	& 0.60	& 0.44	\\
			& graphite 	& ---		& 0.57	& $5.3 \times 10^3$	& 0.61	& 0.47	& 0.24	\\
\tableline
\end{tabular}
\tablenotetext{a}{See Section 5.1 and Table \ref{tbl_dust_properties} for details.}
\tablenotetext{b}{We adopt the mass density 2.26 g cm$^{-3}$ for graphite \citep{Kim94}, and 1.44/1.67/1.84/1.99 g cm$^{-3}$ for amorphous carbon with pyrolysis temperature of 400/600/800/1000 $^\circ$C, respectively \citep{Jager98}.}
\end{center}
\end{table*}


\clearpage

\begin{deluxetable}{lrccccc} 
\tablecolumns{7} 
\tablecaption{Disk Masses ($M_\sun$)\label{disk_masses}} 
\tablewidth{0pt} 
\tablehead{
\colhead{Dust} &
\colhead{} &
\multicolumn{2}{c}{$\tau = 1$ at $\theta=30^\circ$} &
\colhead{} &
\multicolumn{2}{c}{$\tau = 1$ at $\theta=25^\circ$} \\
\cline{3-4} \cline{6-7} \\
\colhead{} &
\colhead{$\beta$} &
\colhead{$h_{50 \rm AU}$=5 AU} &
\colhead{25 AU} &
\colhead{} &
\colhead{5 AU} &
\colhead{25 AU}
}
\startdata 
KMH&1.3&
$1 \times 10^{2}$&---&& 0.5 &---\\
&2.0&
2 &$1 \times 10^{-4}$&&$5 \times 10^{-2}$&$6 \times 10^{-5}$\\
&2.7&
0.1&$2 \times 10^{-4}$&&$8 \times 10^{-3}$&$1 \times 10^{-4}$
\vspace{0.1cm}\\
C01&1.3&
$6 \times 10^{1}$&---&&0.3&---\\
&2.0&
1&$5 \times 10^{-5}$&&$2 \times 10^{-2}$&$3 \times 10^{-5}$\\
&2.7&
$6 \times 10^{-2}$&$8 \times 10^{-5}$&&$4 \times 10^{-3}$&$5 \times 10^{-5}$
\vspace{0.1cm}\\
C01$\times$15&1.3&
$1 \times 10^{2}$&---&&$5 \times 10^{-1}$&---\\
&2.0&
2 &$1 \times 10^{-4}$&&$5 \times 10^{-2}$&$6 \times 10^{-5}$\\
&2.7&
0.1&$2 \times 10^{-4}$&&$8 \times 10^{-3}$&$1 \times 10^{-4}$
\enddata 
\end{deluxetable} 


\begin{deluxetable}{lrccccc} 
\tablecolumns{7} 
\tablecaption{Visual extinction ($A_V$) toward the star \label{A_V}} 
\tablewidth{0pt} 
\tablehead{
\colhead{Dust} &
\colhead{} &
\multicolumn{2}{c}{$\tau = 1$ at $\theta=30^\circ$} &
\colhead{} &
\multicolumn{2}{c}{$\tau = 1$ at $\theta=25^\circ$} \\
\cline{3-4} \cline{6-7} \\
\colhead{} &
\colhead{$\beta$} &
\colhead{$h_{50 \rm AU}$=5 AU} &
\colhead{25 AU} &
\colhead{} &
\colhead{5 AU} &
\colhead{25 AU}
}
\startdata 
KMH&1.3&
$2 \times 10^{-8}$ & --- && 1.1 & --- \\
&2.0&
$3 \times 10^{-7}$ & 1.7 && 1.9 & 5.1 \\
&2.7&
$5 \times 10^{-6}$ & 2.0 && 2.6 & 5.3\vspace{0.1cm}\\
C01&1.3&
$1 \times 10^{-8}$ & --- && 0.7 & --- \\
&2.0&
$2 \times 10^{-7}$ & 1.0 && 1.1 & 3.0 \\
&2.7&
$3 \times 10^{-6}$ & 1.2 && 1.6 & 3.1\vspace{0.1cm}\\
C01$\times$15&1.3&
$7 \times 10^{-9}$ & --- && 0.4 & --- \\
&2.0&
$1 \times 10^{-7}$ & 0.5 && 0.6 & 1.7 \\
&2.7&
$2 \times 10^{-6}$ & 0.6 && 0.9 & 1.7 
\enddata 
\end{deluxetable} 




\end{document}